%% file: main.tex
\documentclass[11pt,a4paper,copyright,goog]{google}

\usepackage[numbers,compress]{natbib}
\input{macros}

\paperurl{}

\uselogo{} 

\title{Towards Scalable Measurement of Durable Skills}
\author[a]{Amir Globerson}
\author[a]{Amy Keeling}
\author[a]{Anisha Choudhury}
\author[a]{Anna Iurchenko}
\author[d]{Aviad Segal}
\author[a]{Avinatan Hassidim}
\author[a]{Ayça Çakmakli}
\author[a]{Ben Gomes}
\author[a]{Benn Witt}
\author[a]{Cathy Cheung}
\author[c]{Cristine Legare}
\author[a]{Diana Akrong}
\author[d]{Eliad Carmi}
\author[a]{Elisabeth Bauer}
\author[a]{Gal Elidan}
\author[d]{Hadas Gelbart}
\author[a]{Hairong Mu}
\author[a]{Katherine Chou}
\author[a]{Lev Borovoi}
\author[a]{Nir Kerem}
\author[a]{Niv Efron}
\author[a]{Noa Kerrem Gilo}
\author[a]{Preeti Singh}
\author[a]{Rajvi Kapadia}
\author[a]{Rena Levitt}
\author[a]{Roni Rabin}
\author[a]{Ronit Levavi Morad}
\author[a]{Rotem Yulzary}
\author[a]{Shashank Agarwal}
\author[a]{Sophie Allweis}
\author[a]{Tracey Lee-Joe}
\author[a]{Tzvika Stein}
\author[d]{Yael Bar Moshe}
\author[a]{Yael Haramaty}
\author[a]{Yaniv Carmel}
\author[a]{Yishay Mor}
\author[a]{Yoav Bar Sinai}
\author[b]{Yoav Bergner}
\author[a]{Yossi Matias}
\author[a]{Yuri Lev}

\affil[a]{Google Research}
\affil[b]{New York University}
\affil[c]{The University of Texas at Austin}
\affil[d]{OpenMic}

\begin{abstract}
Durable skills, such as collaboration, creativity and critical thinking, are instrumental to success in the modern workforce. Yet, measuring these skills remains a persistent challenge. Moreover, because what is not measured is often not taught, these skills are often overlooked in mainstream educational curricula. 
Designing effective assessments for these skills necessitates balancing two often-conflicting requirements: ecological validity and psychometric rigor. On the one hand, the assessment environment should emulate natural interaction between humans, which is how these skills will be performed in the real world. On the other hand, it should be scalable, controllable and reproducible. Here we argue that Large Language Models (LLMs) can be used to better capture both of these aims. 
Concretely, we develop an AI-based framework where the subject converses with AI teammates in a way that resembles human-human interaction for  authenticity, while also offering the psychometric control required for informative and robust assessment. Importantly, the AI participants not only act as teammates but also, in an ``Executive LLM'' setup, steer the conversation towards eliciting a high density of observable evidence for skill proficiency. We complement this with an AI evaluator that can be used to measure skill proficiency in such interactions. We evaluate our assessment protocol based on transcripts of interactions of human participants with our AI framework, for multiple durable skills. For the skill of creativity, we further demonstrate the efficacy of an autorater for evaluating complex tasks performed by high school students. 
Our analysis shows that the use of the Executive LLM significantly increases
elicited evidence, compared to non-steered
interactions. In addition, we show that LLM-automated scoring
of conversations largely agrees with that of expert annotators.
This research demonstrates the utility of orchestrated LLMs  approaches for measuring complex social and cognitive constructs in a scalable and controllable manner.
\end{abstract}

\begin{document}
\maketitle

\section{Introduction}
Success in the modern workplace requires not only technical knowledge and procedural fluency, but additionally a host of future-ready human skills, including  collaboration, communication, critical thinking, and creativity \cite{burrus2013identifying, darling2013criteria}. Often also referred to as 21\textsuperscript{st} century skills \cite{trilling200921st, voogt2012comparative,hilton2012education, natalie2023innovating, griffin2012assessment}, these constructs remain notoriously hard to measure \cite{stecher2014measuring, liu2016tough, fiore2017collaborative}. 
Previous efforts to resolve these measurement challenges and align educational goals, 
assessment, and instruction have focused on technological innovation, for example through automated scoring of answers \cite{yannakoudakis2011new,taghipour2016neural,alikaniotis2016automatic} and the collection and analysis of detailed process data \cite{bergner2019process,piech2015deep}. 
Progress in large language models (LLMs) has opened up a new technological frontier, which we pursue here. 
Our key idea is that LLMs can bridge the gap between unstructured student collaboration, which more closely emulates classroom practice, and standardized assessment, which, while artificial, attempts to isolate the behaviors needed for valid inference. 

\begin{figure*}[t!]
    \centering    
    \includegraphics[width=1.0\linewidth]
    {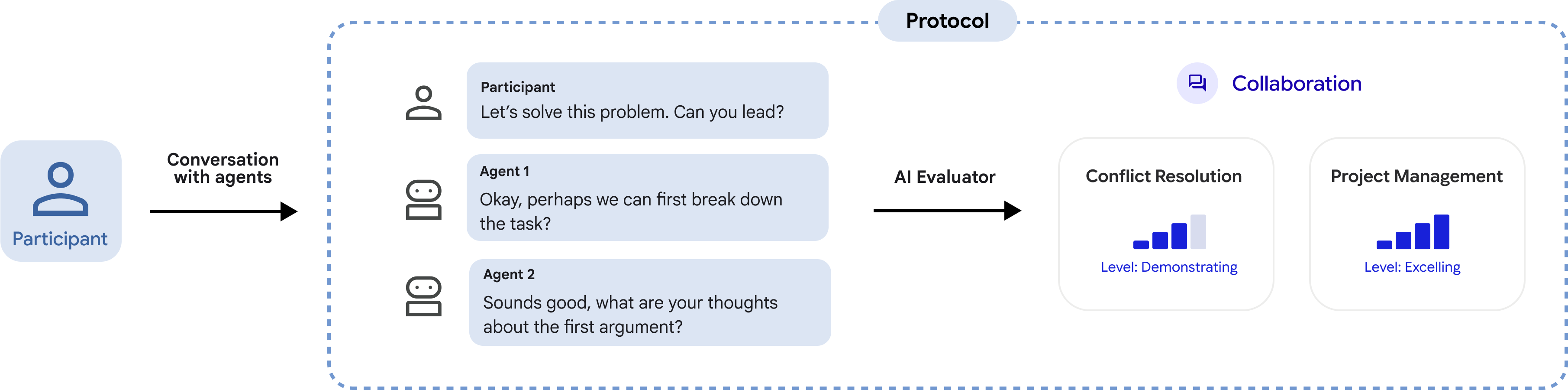}
    \caption{An illustration of the AI-based assessment protocol. A human participant engages in conversation with AI-based teammates to solve a given task (e.g., collaboration in this example). After the conversation concludes, an \AIscorer LLM analyzes the transcript and provides an assessment of skill levels along several dimensions (in this case Conflict Resolution and Project Management).
    }
    \label{fig:protocol_with_human_subject}
\end{figure*}

Group work is a key concept in education, and is viewed as beneficial to individual learning itself \cite{dillenbourg2009evolution} and to the development of durable skills \cite{johnson2000cooperative}. However, assessing the individual in such settings is especially challenging from a psychometric perspective due to the inherent interdependence between interacting members. Only a handful of large-scale efforts have addressed this challenge, the most notable being the Assessment and Teaching of 21\textsuperscript{st} Century Skills project (ATC21S)\cite{griffin2012assessment} and the PISA 2015 CPS assessment \cite{nouri2017assessing, he2017collaborative, stadler2020assessment}. 
While each program defined its own framework for the measurement construct, the differences in theoretical approaches were relatively minor. More profound were the different operationalizations of the construct. PISA employed an environment where the human subject interacted with scripted simulated teammates via multiple choice questions. In contrast, in the ATC21S project assessment tasks were carried out in human-human dyads, solving collaborative tasks by acting on objects in a digital environment (e.g., passing objects to one another) as well as communicating via a chat window. These approaches demonstrate two points on the spectrum between naturalistic and fully structured interactions, yet both are still far from true group interaction observed in classrooms. 

Motivated by the above, we ask whether it is possible to develop an assessment approach that captures the open-ended nature of human-to-human interaction, while avoiding the variance of unstructured ``in-the-wild'' interactions. We argue that LLMs (e.g., \cite{team2023gemini,gpt4}) offer an effective approach to the problem. Prior to the availability of LLMs, incorporating artificial personas in a simulation for assessment required mostly hard-coded rules, thus constraining the simulated group work. This was intentionally the case for the human-agent tasks in large-scale assessments (e.g., PISA 2015 \cite{stadler2020assessment}).

Naturalistic conversational interactions are a native affordance of LLMs, opening the door to less structured, more authentic interactions. However, reduced structure often reduces the guarantee of observing the specific behaviors necessary for inference. Indeed, \citet{sijtsma2011introduction} observed that measurement is a compromise in the name of efficiency since the ``long lasting observation of a person in real life until (s)he spontaneously exhibits the behavior of interest... would take too much time before enough evidence was collected.''  Here we address this challenge by introducing the \puppet, designed to explicitly drive the conversation towards eliciting evidence of specific skills and maximizing assessment accuracy, while keeping the conversation natural. Our approach can therefore be viewed as an adaptive test of complex behaviors. 

We implement the \puppet within a scalable virtual assessment experience called \skillup, which resembles the flexible nature of human-human interactions while offering sufficient structure for robust assessment of future-ready skills. In \skillup, a human subject interacts with one or more AI-based teammates to carry out a joint task. 
The group tasks are designed to resemble authentic classroom tasks, and a rubric is developed to evaluate the human participant in the virtual interactions along dimensions relevant to the skill of interest. The {\puppet} generates the responses for all the AI teammates in the conversation and is designed to steer the conversion toward maximal information and assessment accuracy.
Finally, a separate LLM is used to assess the transcripts of the human-AI interactions. See \figref{fig:protocol_with_human_subject} for an illustration of the overall assessment protocol.

The above protocol was used for assessing the skills of collaboration, creativity and critical thinking, each in the context of group tasks appropriate for the elicitation of evidence for that skill. In order to validate the assessment approach, we focused on the complex skill of collaboration, and collected conversations of human participants interacting with \skillup as well as ratings of these conversations by expert pedagogical raters. We additionally analyzed simulated conversations for the skills of creativity and critical thinking.


We posited several key questions for our assessment approach. First, can the skills of participants be reliably scored by human raters based on transcripts from the conversations generated in \skillup? Second, can the {\puppet} improve evidence accumulation by steering these human-AI conversations? Third, can automated scoring using an \AIscorer agree with human raters as consistently as human raters agree with each other? Finally, we ask whether such evaluation protocols can be refined by simulating human subjects. Our results provide positive evidence for all four questions, thus establishing the {\puppet} approach as a promising mechanism for assessing proficiency in complex durable skills. Finally, for the skill of creativity, we also show that for complex creativity tasks submitted by high-school students, a Gemini based AI Evaluator is an effective creativity assessor, on par with human expert raters. 

\section{A Scalable Assessment Protocol for Human Skills}
We consider the setting where a human subject is to be evaluated on a skill $S$, and a multidimensional scoring rubric is given for that skill. In our setting, $S$ will correspond to collaboration, creativity, and critical thinking. The outcome of the assessment is a rating of the subject on each of the dimensions of skill $S$ according to the rubric. We propose to measure the skill level of the subject by having them converse with AI teammates. The human and AI teammates are given a task they need to solve together.
We now briefly discuss the components involved in this setting, illustrated in \figref{fig:protocol_with_human_subject}.

\subsection{The {\puppet}}
Perhaps the simplest approach to constructing a group of LLM teammates is to literally construct a group of LLMs, and allow these to interact with the human and with each other with no further constraints. However, as we show in the experimental evaluation, this ``{\nonpuppet}'' setting does not result in sufficiently informative interactions and more control is needed to elicit interactions that can be robustly assessed for skill proficiency.

\begin{figure*}[!t]
    \centering
    \includegraphics[width=0.95\linewidth]{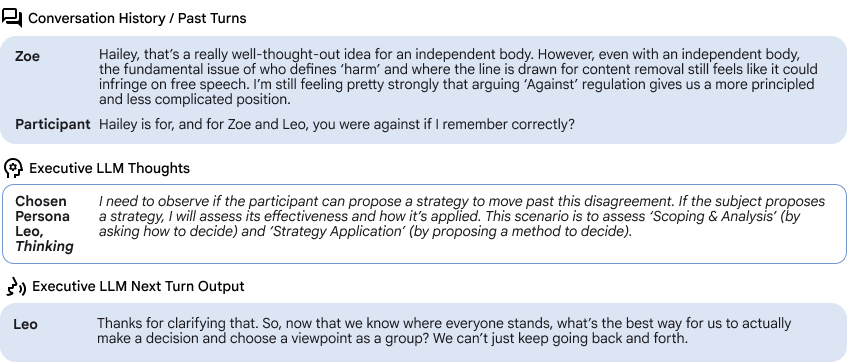}  \\
    \caption{An excerpt from a  conversation between a human participant and the {\puppet}. The task is Debate and the {\puppet} focus is on conflict resolution. The {\puppet} thinks about the participant's response and chooses to respond as "Leo". Also shown is the internal reasoning  of {\puppet} before generating the final response as "Leo".}
    \label{fig:sample_conv}
\end{figure*}

Instead of having separate LLMs model each of the AI teammates, we employ a single LLM which generates the responses for all of them. This {\puppet} has access to pedagogical rubrics that will also be used for assessment, and is prompted to generate text that will maximize the information and accuracy of the assessment protocol. Namely, it should create situations that will allow the subject to demonstrate their skill, so that their skill level can be quantitatively inferred. As an example, if ``Conflict Resolution'' is evaluated, the {\puppet} may have one of the AI teammates start a conflict and maintain it until evidence of conflict resolution skill has been observed. \figref{fig:sample_conv} shows a sample excerpt from a conversation in the {\puppet} protocol, where the {\puppet} steers the conversation towards eliciting more information about the participant's skill-related behavior. 

\subsection{Automatic Assessment and Feedback}
Given a transcript of the interaction between the subject and the AI teammates for a task $T$ and skill $S$ and the corresponding evaluation multi-dimensional rubric (see below), our goal is to evaluate the level of the subject on each of the rubric dimensions. For each of the subject's turns in the conversation, we use an LLM (Gemini 3.0) to generate ratings at a turn level. The LLM was provided the conversation and prompted to output
a level for each of the skill dimensions per turn. It also had the option to return NA, indicating no evidence for the given dimension in the turn.
This rating was repeated 20 times per turn, and the final turn label was generated from these values, as follows. The final turn label was declared to be NA if at least one of the 20 predictions for the turn was NA. The final turn level rating (if not NA) was the most frequent level among the 20 returned.
Then, to obtain a rating for the entire conversation, we took the turn outputs and trained a regression model (linear regression for the scores and logistic regression for the NA) based on human ratings. Model performance was evaluated using leave-one-out cross-validation. 

\begin{figure*}[!t]
    \centering
    \begin{tabular}{c}
    \hspace{-1in} \includegraphics[width=0.7\linewidth]{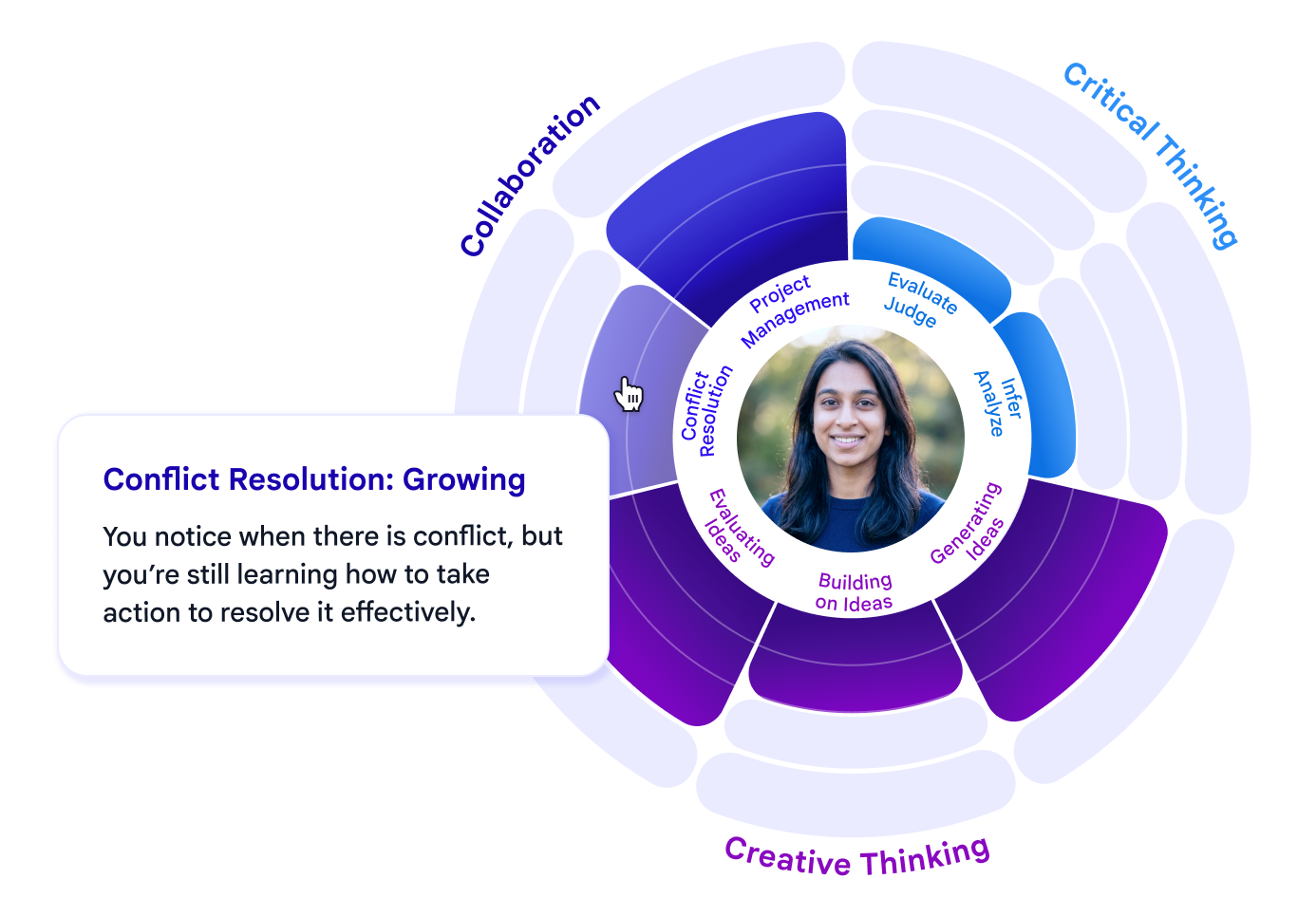} \\
    \includegraphics[width=0.9\linewidth]{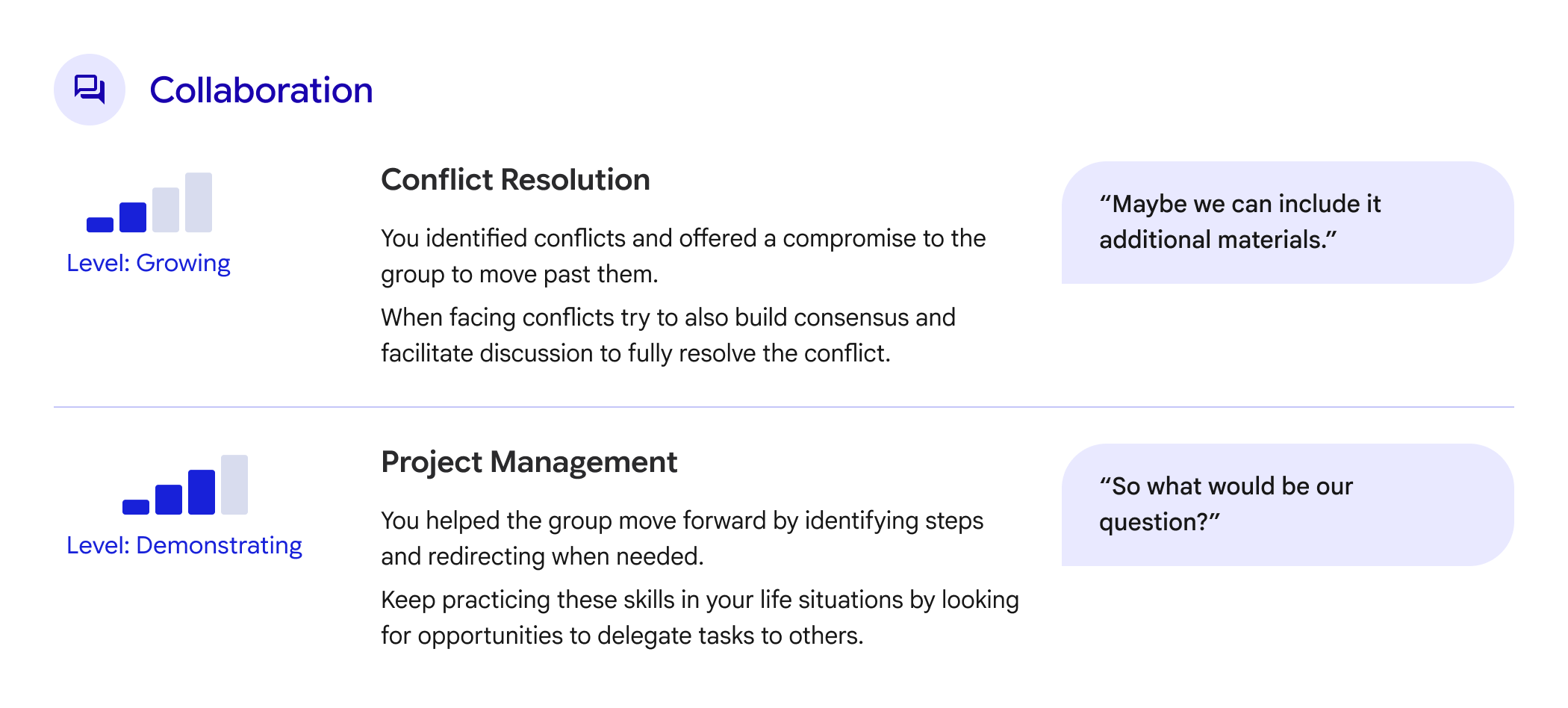} 
    \end{tabular}
    \caption{The assessment and feedback provided to the user in \skillup. The skills map (top) shows quantitative competency of all skills practiced, along with a breakdown into the axes evaluated for each skill. Each can be expanded to also get qualitative feedback that explains the rating. The user can further drill down to a more detailed view (bottom) that provides excerpts from the conversation that substantiate the analysis.}
    \label{fig:skillmap}
\end{figure*}

Feedback is provided to the user in \skillup in a quantitative manner, along with qualitative evidence taken from the interaction with the AI teammates. An overall map gives a broad view of competency for all skills, with a break-down for different axes or sub-skills for each one. The user can further drill down and receive excerpts from the conversation which exemplify the numeric competency score. See \figref{fig:skillmap}. 

\subsection{Skill Rubrics and Tasks} 
The \skillup system focuses on the core durable skills of collaboration, creativity and critical thinking. For each of those, we developed scoring rubrics, based on previous work. First, the relevant literature was reviewed in order to construct a conceptual model of the skill. From this, an initial rubric was derived. This rubric was used by human experts to assess sample conversations, and refined where agreement was low or the raters found it ambiguous. This approach is consistent with \cite{haocui2025}, who also found that LLMs are reliable for coding conversations when given a rubric that is derived from theory and then refined via expert use on sample data. Each rubric consisted of multiple dimensions, each with a score of 1-4, and an NA option, signifying there was insufficient information to decide on a score. See below for the conceptual basis of rubrics for each skill, and the appendix for the complete rubrics. The rubrics were provided as input to both the \puppet and the \AIscorer. 

\begin{figure*}[!t]
    \centering
    \includegraphics[width=1.0\linewidth]{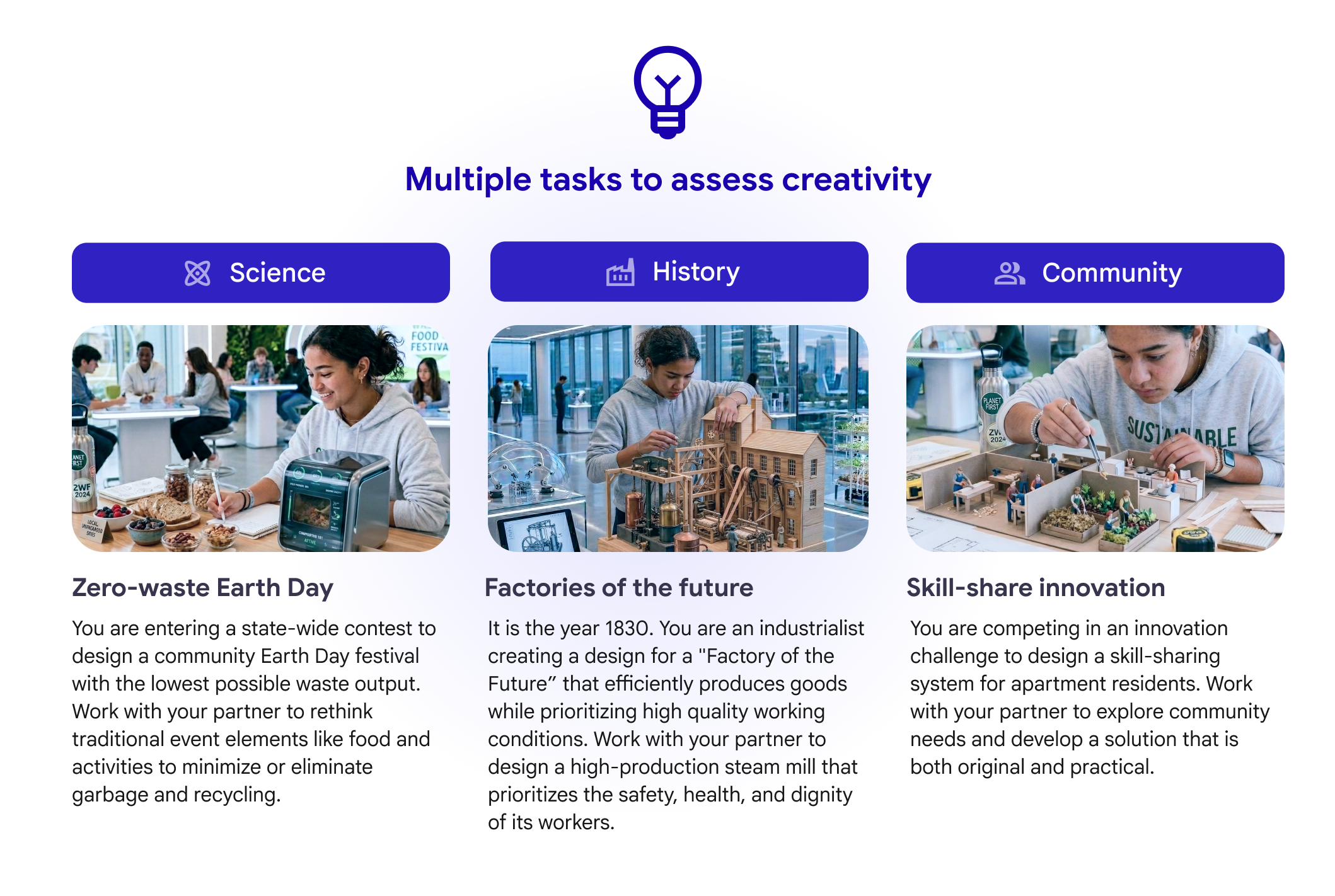}  \\
    \caption{Example tasks in \skillup for assessing and practicing the skill of creativity.}
    \label{fig:projects}
\end{figure*}

For assessment and practice in \skillup, our pedagogical experts chose several themes for projects similar to those encountered in real classroom. For each theme, specific tasks for assessing and practicing the skills of collaboration, creativity and critical thinking were designed. See \figref{fig:projects} for a short description of the tasks for the creativity skill in \skillup. See the appendix for the conceptual basis of the tasks and rubrics for all skills.

\section{\skillup Experimental Setup}

In our \skillup system a human subject interacts with virtual AI personas to complete a task aimed at eliciting sufficient information for skill assessment. We now describe the setup used for the evaluation of the \skillup skill assessment protocol. For evaluation, we focus mostly on the skill of collaboration where the user interacts with multiple AI teammates. We also present initial results for the creativity and critical thinking skills.

\subsection{Environment}
Participants who took part in the study were told they would be engaging in collaborative group activities with a set of AI teammates. Participants accessed a user interface that introduced them to their group of AI teammates and described the task they were to complete. Communication took place via a chat-like user interface which allowed interacting with the teammates. Each AI teammate had a name, and their contributions to the chat were delivered via both text and audio. Participants had the option to contribute to the discussion by either typing their responses or using a voice interface. Following the task description, they were requested to interact with the teammates for 30 minutes per conversation. Gemini 2.5 Pro \cite{Gemini25} was used as model underlying the \puppet, that generated responses for all AI participants in the main body of experiments for the skill of collaboration. For the additional initial results for creativity and critical thinking in \secref{sec:res_other_skills}, Gemini 3 was used.

\subsection{Participant recruitment}
Participants ages 18-25 were recruited via the Prolific platform. They were English native speakers based in the United States. 
188 participants were recruited, and 
each generated two conversations, resulting in a total of 373 conversations (three conversations were filtered due to technical issues). 

\subsection{Experiment design} 
In order to check the efficacy of the {\puppet} in steering skill-related behavior, we collected conversations driven by an {\puppet}, and also with {\nonpuppet}. Specifically, we considered an {\puppet} that is prompted to focus on assessment of one of the sub-dimensions of collaboration: Conflict Resolution or Project Management. 
For each of the tasks, the protocol was randomized to be either an {\puppet} focusing on Conflict Resolution or on Project Management or to use {\nonpuppet}. We also tested whether telling the subject to focus on the skill had an effect. Namely, before the conversation started we added the text: ``In the following task, you are asked to pay particular attention to your role in [skill].'' where skill was either ``Conflict Resolution'' or ``Project Management''. We randomly assigned subjects to receive either no focus instructions (50\%) or instructions focusing on one of the two skills (25\% each).

\subsection{Human rating of conversations}
Two pedagogical raters from New York University were recruited to annotate the conversations. They were provided with a scoring rubric for conflict resolution (CR) and for project management (PM). Raters were asked to consider each human turn of the conversation and, first, decide whether it should count as evidence of CR, or PM, or not. If not, they selected NA from a pull-down menu. If so, they selected a particular component of CR or PM from the rubric and provided a score on a scale of 1-4. In addition to the turn by turn rating, raters were also asked to provide a holistic score for the conversation. This holistic score could again be NA, if there was not enough evidence to decide, or a number between 1-4.

\subsection{Simulated Conversations} 
Since evaluation with human participants is costly, it is worthwhile to have a sandbox that can simulate human participants, towards developing the protocol before deployment. Towards this end,
 Gemini was used to simulate the human participant in the conversation. 
 For the recovery evaluation described in \secref{subsec:sim_colab}, Gemini was prompted to behave like a student evaluated on their collaboration skill, and was also instructed to behave according to a specific level of one of the rubric dimensions (e.g., level 3 of conflict resolution). This skill level was later compared to the level returned by the \AIscorer. Each conversation contained 50 turns, and each level was repeated 100 times. We also used simulation to assess evidence level (e.g., Figures \ref{fig:not_na_cr_ct} and \ref{fig:sim_not_na}), and in these cases the simulated subject was not provided with a skill level.

\section{Results}
We start by assessing the validity of our assessment approach, evaluating the quality of the \AIscorer as well as the efficacy of the {\puppet}, using both interactions with human participants as well as simulated subjects. These results are reported for the collaboration skill in \secref{sec:collab_results}. We then provide additional results on other skills in \secref{sec:res_other_skills}. Finally, we demonstrate the efficacy of our assessment autorater in a real student setting for a complex creativity task in \secref{sec:openmic}.

\subsection{Evaluation of the \skillup assessment protocol for Collaboration
\label{sec:collab_results}}
We carry out a comprehensive evaluation of our approach in the context of the collaboration skill, where the group setting is the most challenging and involves four members -- the human participant and three AI team members.

\subsubsection{LLM rating is on par with expert human raters}
\begin{figure*}[t!]
    \centering
    \includegraphics[width=0.7\linewidth]{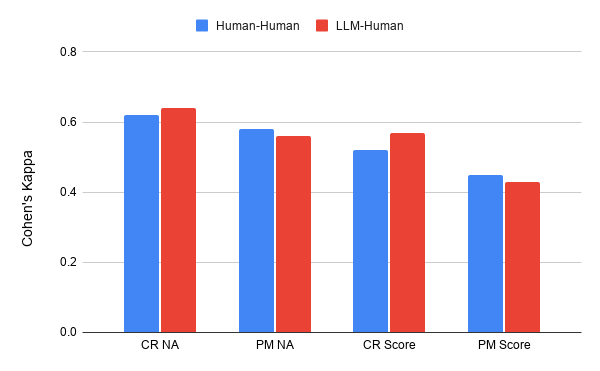}  \caption{Conversation level agreement using Cohen's Kappa ($\kappa$) for binary outcomes (NA or not), and quadratically weighted Kappa for scores, for conflict resolution (CR) and project management (PM). Compared are the inter-expert agreement (blue) and the LLM-expert agreement (red). 
    \label{fig:projects_alternative}}
\end{figure*}

Roughly half of the conversations received a single human expert rating, and half of those also received an additional human expert rating. This allowed for a measure of inter-rater agreement between the two human experts and between the LLM \AIscorer and the human experts. Agreement statistics using Cohen's Kappa \cite{Kappa, WeightedKappa} for conflict resolution (CR) and project management (PM) are shown in Figure \ref{fig:projects_alternative}. These include score-or-no-score (i.e., NA) decisions, as well as finally numerical score agreements (for cases where both raters provided a numerical score). 


Coding conversations for conflict resolution and project management proved to be challenging, even after several rounds of rater calibration. Agreement between the human expert raters was moderate, with Kappa in the range of 0.45-0.64 \cite{Landis+Koch:1977}. Agreement between the LLM and the human experts was similar to that between the two experts. This suggests that automatic rating of conversations is a scalable alternative to human rating in our assessment setup.
Having established that the LLM-based autoraters provide similar results to those of human raters, below we use only autoraters, which can be run at a larger scale, to substantiate the validity of the assessment protocol. Results with human ratings are qualitatively similar to those of the autoraters.

\begin{figure}[t]
    \centering
    \includegraphics[width=0.95\linewidth]{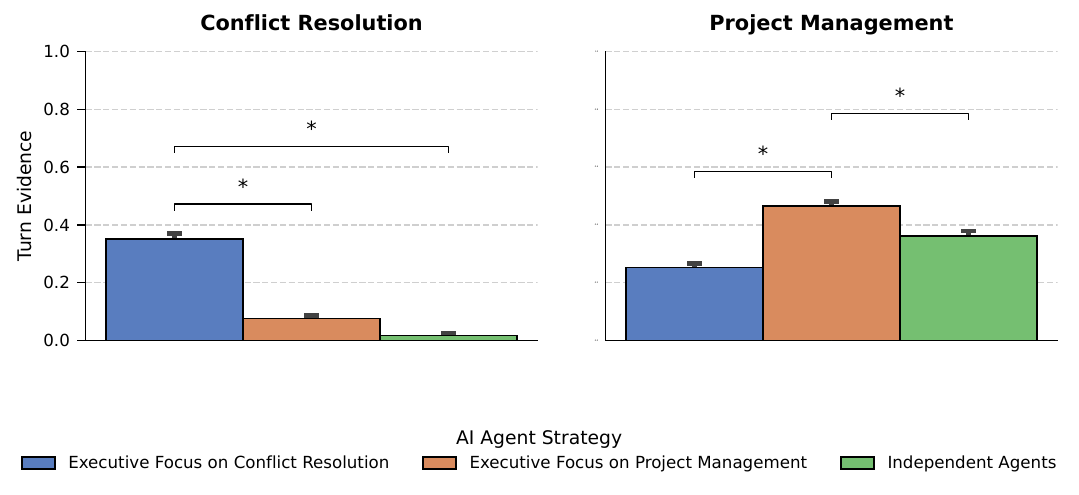}  \\
    \caption{Turn-level skill evidence. The fraction of participant turns rated as demonstrating a given skill for conflict resolution and project management. Results are shown for two versions of an {\puppet}, each focusing on a different skill, and the {\nonpuppet} protocol. Starred brackets denote a statistically significant difference ($p\leq 0.05)$ using the Fisher exact test.
    }
    \label{fig:steerability-turn}
\end{figure}

\begin{figure}[tbhp]
    \centering
    \includegraphics[width=0.95\linewidth]{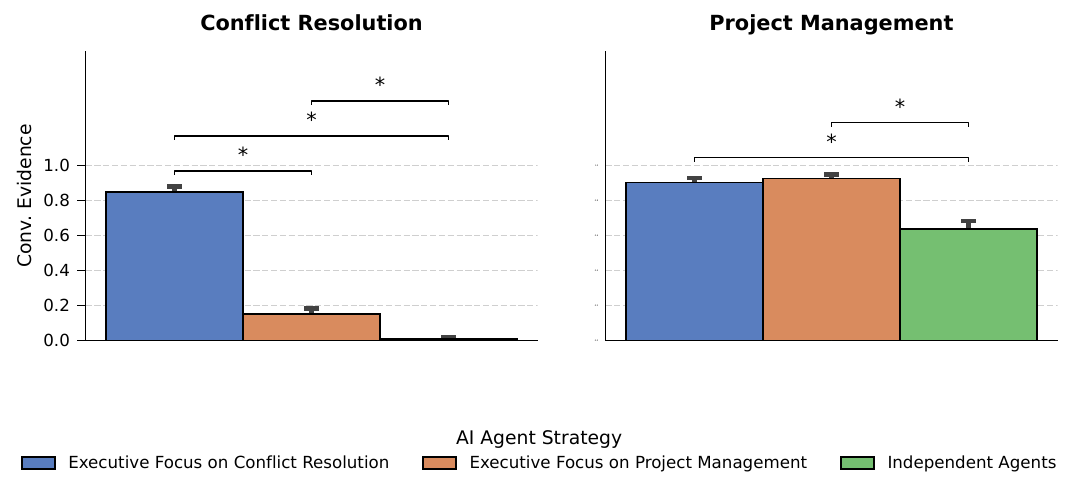} 
    \caption{Conversation-level skill evidence for the collaboration skill. The fraction of conversations that were rated as having sufficient information to provide a skill-level rating (i.e., not rated as NA). Results shown for the Conflict Resolution (Left) and Project Management (Right) sub-skills.
    Results are shown for two versions of an {\puppet}, each focusing on a different skill, and the {\nonpuppet} protocol. Starred brackets denote a statistically significant difference ($p\leq 0.05)$ using the Fisher exact test.}
    \label{fig:steerability-conversation}
\end{figure}
\subsubsection{{\puppet}s Elicit Skill-Related Information}
The goal of the assessment task design is to elicit evidence of the skill of interest. A conversation-based group task \textit{can} create opportunities for the subject to demonstrate conflict resolution (or project management) skills. But such evidence need not necessarily arise, for example if the group happens to work well and with no friction. We hypothesized that conversations using the {\puppet} protocol would do a better job of eliciting evidence of the target collaboration skills. If so, this effect would be observable both at the turn-level and at the conversation-level, though not necessarily in the same quantity.
The measure of evidence elicitation at both levels was simply calculated as the fraction of turns or conversations that were coded as evidence for the skill as opposed to as NA.

\figref{fig:steerability-turn} (turn-level) and \figref{fig:steerability-conversation} (conversation-level) show evidence levels for Conflict Resolution (CR) and Project Management (PM) in each of three different AI agent protocols: a CR {\puppet}, a PM {\puppet} and {\nonpuppet}. 
First, it can be seen that an {\puppet} focused on a skill always elicits significantly more evidence for that skill than the {\nonpuppet}.
Second, there is a desirable cross-over effect between the two versions of the {\puppet} when matched or mismatched to the measured skill. That is, steering the conversation towards CR increases evidence of CR but reduces evidence of PM and vice versa. This results in a significant difference in three of the four figures, except for measuring PM at the conversation level (\figref{fig:steerability-conversation}, Right), where similar information levels are obtained for both {\puppet} versions. Finally, the conversation-level information rate is quite high, at $92.4\%$ for PM and $85\%$ for CR, when the skill-matched {\puppet} was used. Taken together, these results indicate that use of {\puppet} extracts more informative interactions with respect to the skill that the {\puppet} is focused on.

It is also interesting to note that evidence rates are generally higher for Project Management than for Conflict Resolution, for almost all cases (except Executive Focus on PM at the turn level). This agrees with the intuition that project management behaviors are more abundant, and also require less steering than conflict resolution.

\subsubsection{The Effect of Task Topic - Science vs Debate} 
 The experimental setup involved two distinct tasks that subjects solved with the AI teammates: science and debate. To determine whether the type of task had an effect on conversation informativeness, independent of the influence of the Executive LLM, a logistic regression analysis was conducted. Separate models were fitted to predict the binary informativeness scores for both skills. To evaluate the impact of the task type, the coefficient for the task variable in each model was tested against the null hypothesis that it was equal to zero. The results revealed no statistically significant difference from zero for either metric (p = 0.18 for Conflict Resolution informativeness; p = 0.9 for Project Management informativeness). Consequently, the analysis establishes that the assigned task does not significantly alter the informativeness of the conversation. This demonstrates the potential for using the the same framework on varied tasks and subject areas.
 
\subsubsection{Is Telling the Subject Sufficient Without Steering?} The {\puppet} is meant to elicit skill-related behavior from the subject. We checked if such elicitation can be obtained by simply asking the subject to focus on a skill. To test this, subjects were randomly selected to either receive focus instructions or not. The focus instructions told the subject to focus on a particular skill. When the protocol was {\puppet} that skill was the same skill that the {\puppet} was focused on. Otherwise it was randomly chosen. To test the statistical effect of the focus on conversation informativeness, we used Fisher's exact test to check whether setting subject focus to a skill improved conversation information  levels vs the no focus setting. Results showed that subject
focus had no significant effect, both in the {\nonpuppet} and {\puppet} settings and for both skills (all at $p>0.6$).
This suggests that the high information rates observed
for {\nonpuppet} cannot be achieved via subject focus instructions alone.

\begin{figure}
    \centering
    \includegraphics[width=1.0\linewidth]{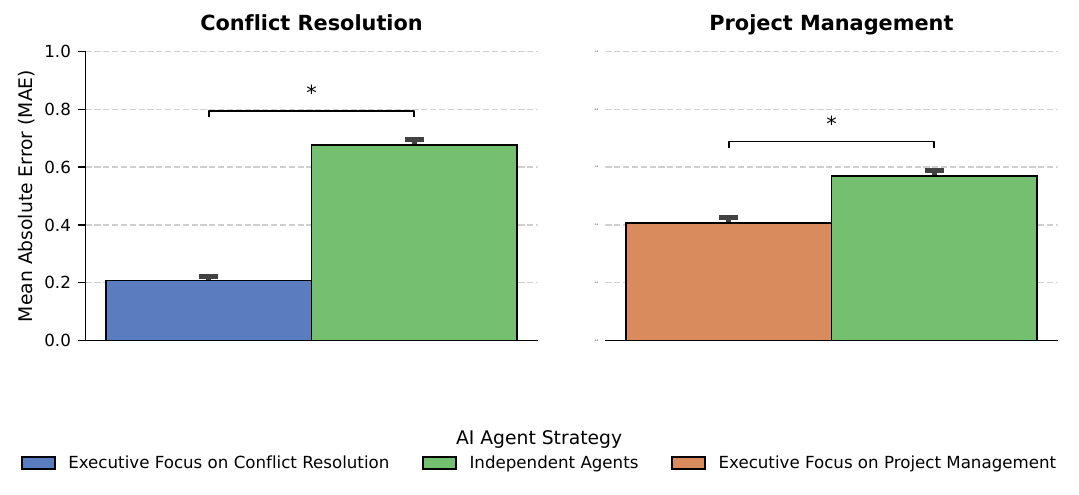}
    \caption{Recovery error for different protocols. An LLM is prompted to simulate a subject at a given skill level $L$. The simulated subject then goes through the assessment procedure and the conversation is mapped to a skill level $\hat{L}$ using an autorater. The reported MAE is the mean absolute difference $|\hat{L}-L|$. Results are reported for both the conflict resolution and project management skills, comparing the {\nonpuppet} and the corresponding {\puppet} protocol for each. Starred brackets denote a statistically significant difference using the Student's t-test.}
    \label{fig:alignment_mae}
\end{figure}

\begin{figure}
    \centering
    \includegraphics[width=1.0\linewidth]{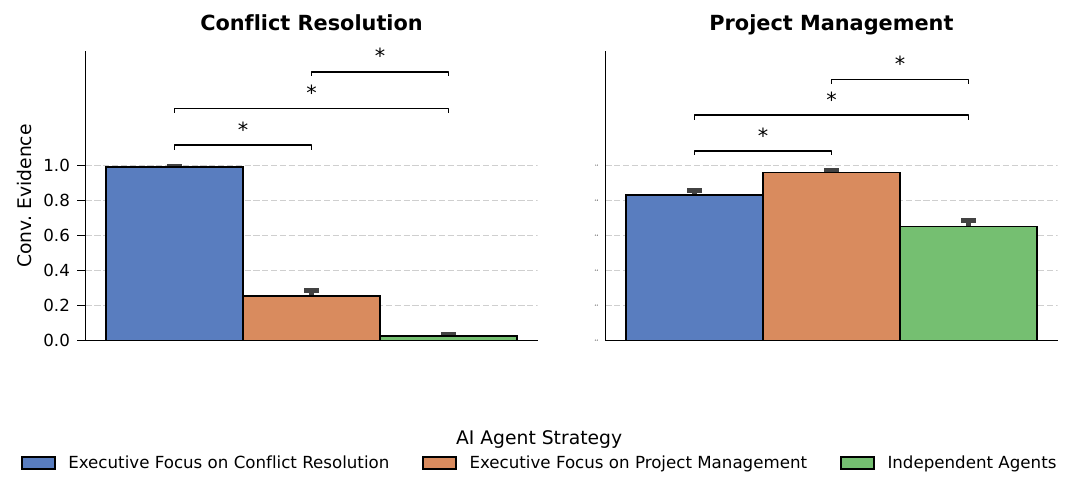}  \\
    \caption{Conversation-level skill evidence for simulated conversations. The setting is as in \figref{fig:steerability-conversation}, but the subject is simulated using an LLM. Starred brackets denote a statistically significant difference using the Fisher exact test.
    }
    \label{fig:sim_not_na}
\end{figure}

\subsubsection{{\puppet}s Improve Skill Assessment - \\ Recovery of ``Known" Skill Proficiency of Simulated Subjects}
\label{subsec:sim_colab}

Simulation studies (sometimes called Monte Carlo studies) in the psychometric literature are commonly used to evaluate new methods of inference \cite{feinberg2016conducting}. Using item response theory, for example, it is straightforward to generate response probabilities from a latent trait and its distribution. Recovering the parameters used to generate the data, especially when complexity arises due to mixtures of populations, data censorship, or other confounds, is one step in validating the inferential capabilities of the model. Less commonly, process data such as sequences of steps or actions may also be generated from a sequence model, such as a Markov model (e.g., \cite{tang2020latent}), and then used again to test the inference methodology. Generative AI and LLMs constitute a relatively recent, non-parametric approach to both generating simulated data and inferring traits.

In the present study, we used the conflict resolution and project management rubrics as prompt inputs to a simulated collaborator. This was effectively another agent but one who is ``external'' to the task, as opposed to the task team {\nonpuppet} and/or {\puppet}. We start by defining the ``ground-truth'' conflict resolution or project management skill level of the subject, each on a scale of 1-4, and then run the task with the simulated subject $100$ times for each version of the protocol. We then rate each conversation using the same autoraters for CR and PM as used for the real data.

Figure \ref{fig:alignment_mae} reports recovery results for the different assessment protocols. Recovery error is measured as the mean absolute difference (MAE) between the ``true'' and inferred skill scores. It can be seen that {\puppet} results in overall improved recovery rates, relative to {\nonpuppet}. In addition, we can use the simulated conversations to calculate the same evidence rate metrics calculated on human conversations. Figure \ref{fig:sim_not_na} reports information level in the simulated conversations. Comparison of Figures \ref{fig:steerability-conversation} and \ref{fig:sim_not_na} shows that results are qualitatively similar between the human and simulated conversations. As might be expected, fully simulated conversations are a bit too ideal in that with the corresponding {\puppet}, the conversation evidence rate approaches 100\%. Taken together, the results suggest that rubric-based LLM simulation can be a sandbox for developing improved assessment protocols, reducing some of the costly collection of data with human subjects. 

\subsection{Evidence Rates for Critical Thinking and Creativity
\label{sec:res_other_skills}}



\begin{figure}
    \centering
    \includegraphics[width=1.0\linewidth]{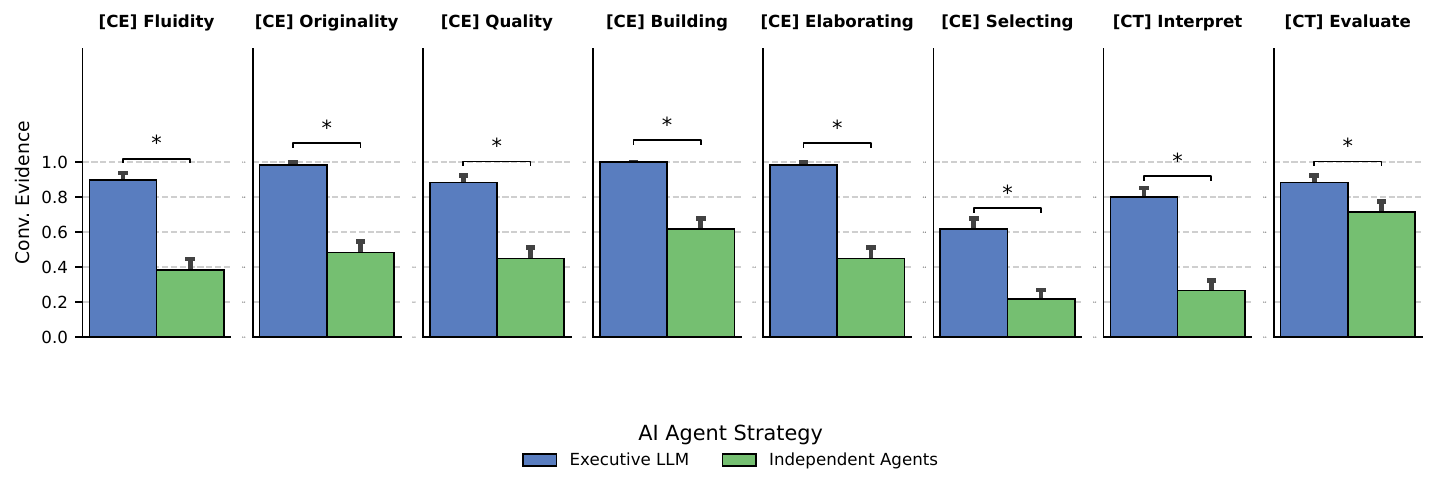} 
    \caption{Conversation-level skill evidence for the different axes of Creativity (CE) and Critical Thinking (CT), for simulated conversations. The setting is as in \figref{fig:steerability-conversation} but for each skill (CE and CT) we use one {\puppet} focused on all dimensions of this skill, and the evidence rate for that {\puppet} is reported. For Critical Thinking the full names of the dimensions are ``Interpret and Analyze'' and ``Evaluate and Judge'', and for Creativity ``Building'' is short for ``Building on Ideas''.}
    \label{fig:not_na_cr_ct}
\end{figure}

Above we presented detailed results for the skill of Collaboration, where we collected human ratings, and compared those to LLM-based ratings. For critical thinking and creativity, collection of human ratings is ongoing, and results will be shared at a later time. Here we present initial results for these skills, on conversations where the subject is simulated (as in \secref{subsec:sim_colab} for the skill of collaboration).

Figure \ref{fig:not_na_cr_ct} shows conversation-level evidence rates for the different dimensions of the creativity and critical thinking skills, when using either {\puppet} or {\nonpuppet} as the AI participants.\footnote{The NA conversation-level rating here was obtained by instructing Gemini to output a minimum and maximum score to assign the conversation on a given dimension, and return NA if these were not identical.}
As with collaboration, it can be seen that the use of the {\puppet} achieves conversation evidence rates that are higher that {\nonpuppet}. We note that these ratings still need to be compared to human ratings, but this does suggest that the {\puppet} approach is also effective for the creativity and critical thinking skills.


\begin{figure}[t]
    \centering
    \includegraphics[width=0.575\columnwidth]{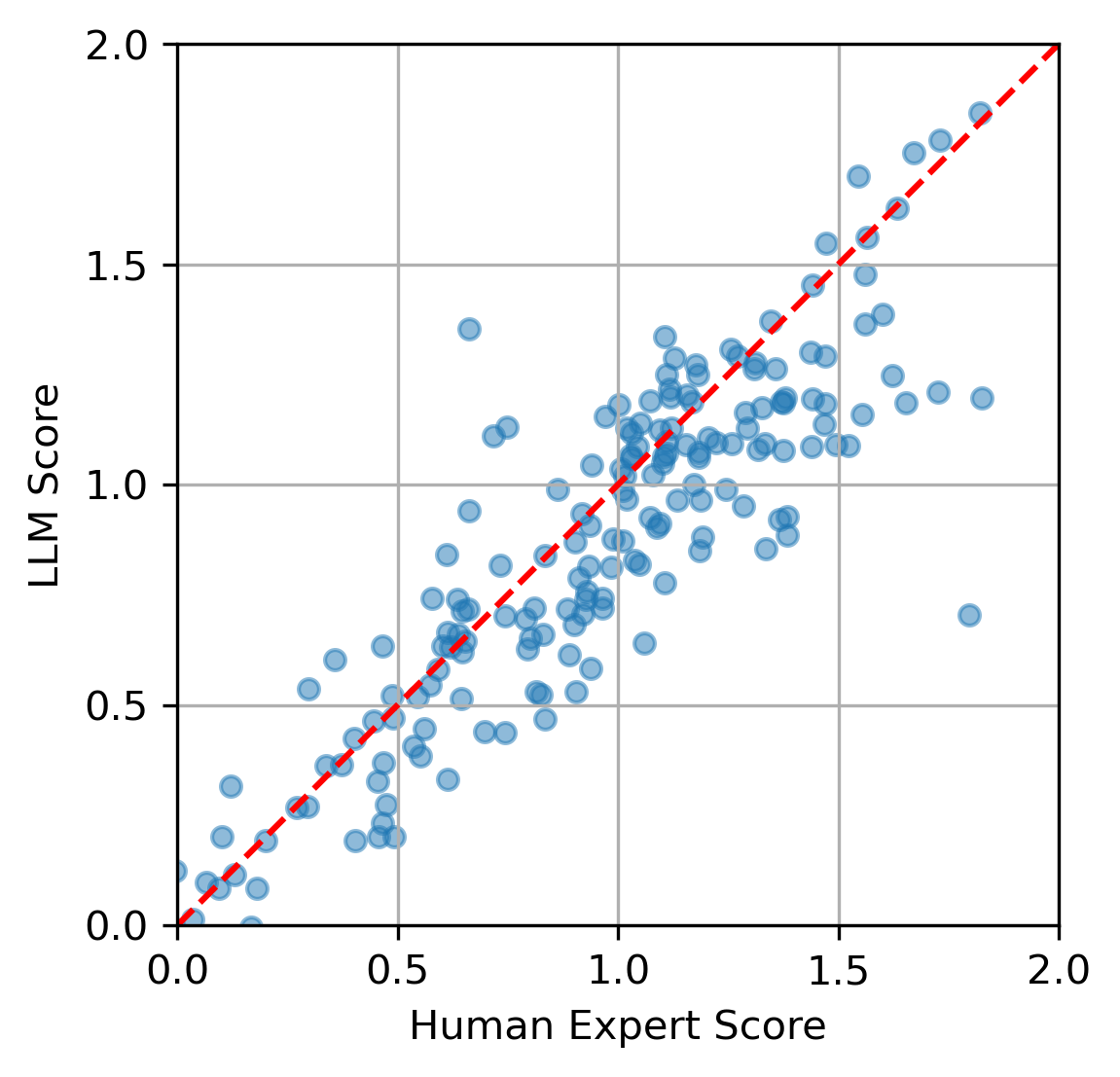}  \\
    \caption{Scatter plot of the autorater scores vs. the human expert scores for held out OpenMic's creativity task submissions. Pearson's correlation between the two is $0.88$.
    }
    \label{fig:OpenMicScatter}
\end{figure}

\subsection{Creativity Assessment in The Real World}
\label{sec:openmic}

In a partnership between Google and OpenMic,
a startup developing AI-powered tools for assessing durable skills, we set out to evaluate the efficacy of our Gemini-based creativity autorater to score complex multimedia creativity tasks given to real students.
Briefly, in a lab study, 280 students were given a short story and were then asked to design a news segment based on the story. This included multiple specific creative tasks such as coming up with character interview questions. The entire experience including the choice of story and specific tasks was designed by OpenMic's creativity experts. These experts also designed detailed rubrics for scoring each item on a scale of 0-2.

The completed tasks were graded by both trained experts and a Gemini-based creativity autorater which took as input the same rubrics used for expert grading. $100$ submissions were used to modify and improve both the Gemini prompt and the expert pedagogical rubrics. The other $180$ submissions were then used to evaluate the accuracy of the autorater relative to the human expert raters.

At the specific item $0-2$ scores, the agreement between the scores of the autorater and those of the human expert raters as measured using Cohen's Kappa was $0.66$, corresponding to good agreement \cite{Landis+Koch:1977}.
More importantly than the per item agreement, a student's proficiency is not measured based on a single item and thus we compare the total grade or sum of the scores of all the items in the submission. A comparison of the autorater overall scores and those of the human experts reveals an extremely high Pearson's correlation of $0.88$. \figref{fig:OpenMicScatter} also shows qualitatively that our automated assessment of creativity for this complex task aligns strongly with those of human experts.

\section{Discussion}
We have argued that large language models have the potential to transform the assessment of complex durable skills. In the case of group work, they help overcome a number of fundamental psychometric challenges regarding the reliability, comparability, scalability, and ecological validity of such assessments. Until now, these challenges have been mostly addressed by highly scripted interactions with AI teammates (e.g., PISA 2015) or highly structured human-human interactions (ATC21S). The novel introduction of the {\puppet} allows standardization of the collaboration experience, without overly scripting the interaction itself.

Group work as defined here is still a human-to-human interaction (although human-agent interaction is rapidly becoming an interesting domain in itself). It follows that the most ecologically valid \cite{schmuckler2001ecological} form of group work assessment would be to put human subjects in groups for collaborative tasks. However, not only is this logistically challenging to scale, but the other team members (relative to any one assessed subject) introduce uncontrolled variance, hence lowering the reliability and comparability of the assessment procedure. Moreover, even if a subject is assessed several times in different groups, in order to counterbalance the variance from other group members, evidence of particular durable skills may or may not manifest naturally. E.g., how can we evaluate conflict resolution if team members all happen to agree?

One way to reduce this variability is by having human actors or ``simulators'' who interact with a subject, following a semi-scripted task. For example, simulation has long been used in medical education \cite{jones2015simulation,Wallace_Rao_Haslam_2002} since the 1960s, and has been shown to be a reliable and effective tool for training and assessment of both core clinical skills and ''soft skills'' such as collaboration and communication \cite{okuda2009utility}. Inspired by this tradition, simulations have expanded into other domains with a clinical character, such as teacher education \cite{kaufman2016enhancing}, as means for training and assessing complex skills. However, scripting tasks with human subjects involves a high degree of tailoring, and is unlikely to be feasible outside of high-stakes contexts. Indeed, since the introduction of LLMs, their use has been explored for simulating human behavior in interaction. Some example applications are training of mental health professionals \cite{elyoseph2026effectiveness}, training teachers \cite{markel2023gpteach}, and medical education \cite{yu2025simulated}.

Replacing human subjects with AI teammates as group members inevitably trades some authenticity in exchange for comparability and reliability. AI teammates powered by LLMs are, at least, much more naturalistic than the rule-based teammates of the past. However, with that freedom comes less control of the conversation. If the conversation is completely unconstrained, then it likely will not be very efficient in providing the evidence of interest about the subject. The {\puppet} presented here addresses this problem by steering the conversation dynamically. 

The {\puppet} approach is analogous to a computerized adaptive test (CAT) designed to increase or decrease the difficulty of test items so as to maximize information about the test-taker \cite{wainer2000computerized}. 
Whereas traditional CAT varies only the difficulty level of items in a narrow domain (e.g., mathematics or verbal skills), the {\puppet} can vary the flow of a conversation in a group task. It does so by modifying the AI participants collectively so as to heighten (or soften) conflicts, create the need for greater project management, or elicit additional creative ideas from the participant.

Our results on skill evidence (Figures \ref{fig:steerability-turn} and \ref{fig:steerability-conversation}) demonstrate that an {\puppet} focused on a given skill can increase the evidence level for the skill of collaboration. This effect is more pronounced in the case of sub-skill of conflict resolution, where the {\puppet} can introduce conflicts explicitly, thus inducing the assessed subject to demonstrate strategies for resolving conflicts. For project management, the {\puppet} still elicits significantly more evidence as compared to {\nonpuppet}, but differences are smaller. This can be explained by the fact that project management behavior arises naturally in conversation, and thus benefits less from steering by the {\puppet}. 

Perhaps the most compelling validity evidence is Criterion Validity \cite{cronbach1955construct}, measuring correspondence between the test results and external measures of performance (either concurrent or predictive). The most reliable version of these would be performance-related metrics such as manager reports, or teacher reports over semesters. However, collecting these is largely impractical due to time and privacy concerns. Thus, we instead propose a measure based on subject simulation, where the test is applied to a simulated subject instead of a human one. Recent results have indeed shown that simulated users can be used to model responses of human subjects \cite{aher2023using,argyle2023out,sarstedt2024using}.
The simulation results we present in \figref{fig:sim_not_na} are in qualitative agreement with those collected from conversation with human subjects in Figure \ref{fig:steerability-conversation}. Furthermore, simulation allows comparison to a ``ground truth'' 
skill-level, and recovery results in \figref{fig:alignment_mae} support the efficacy of {\puppet} as compared to {\nonpuppet}. These results suggest that simulation could be an effective approach towards designing assessment protocols for complex constructs such as collaboration, with improved validity. Finally, we also provided preliminary simulation results for the skills of creativity and critical thinking. The results suggest that the {\puppet} approach is also effective for these skills but further validation with human subjects is needed. We leave this to future work.

Our results for collaboration also demonstrate that LLMs can be used to score conversations according to a rubric, and their agreement with human raters is similar to inter-rater agreement between humans. These results join a growing body of work showing that LLMs can be used to considerably scale the assessment process \cite{kasneci2023chatgpt,ouyang2023artificial,llms_for_feedback}. However, our approach goes beyond rating human-to-human conversations using LLMs, and uses AI for creating conversations with human subjects, in an adaptive manner that provides skill-related information. Finally, we also demonstrated the efficacy of automated assessment for the skill of creativity in the context of a complex creativity task with real students in a lab setting. 

This work focused mostly on developing an AI-based protocol and tool for assessing skills in a group setting. However, AI has great potential for {\em improving} these skills as well. As a simple example, the assessment environment can also be used for continual practice within a low risk sandbox environment, alongside standard group work in the classroom. We leave this possibility for future work. We also recognize that human skills are culturally situated, and will therefore also focus future work on exploring performance across diverse cultural settings and languages to ensure our technology is inclusive and equitable. 
\bibliography{skills}


\newpage
\appendix

\section{Pedagogical basis for Rubrics and Tasks}
We now describe the conceptual basis for the skill specific tasks and associated rubrics, derived from the relevant literature and adapted to our setting. The full rubrics and task descriptions appear in the following appendices.

\subsection{Collaboration}
Our conceptual model of collaboration employs the framework of {\em team cognition}, focusing on the core components of {\em shared mental models} (SMM) and {\em shared situational awareness} (SSA) \cite{cannon-bowerssalas2001, mohammed2010metaphor}. This framework posits that effective collaboration is contingent on a team maintaining a shared understanding of tasks, roles, and contextual beliefs. Team members constantly monitor their interactions (and conversation) to identify and resolve gaps in their SMM or SSA, in a process called {\em grounding} \cite{clark1991grounding}. Two collaboration sub-skills were evaluated: Conflict Resolution and Project Management. For each sub-skill, we used a rubric with several sub-categories, meant to capture different aspects of the skill.  

Following \cite{jehn1995}, our conflict resolution model distinguishes between {\em task conflict} (misalignment on the content of the task), {\em relationship conflict} (interpersonal frictions), and {\em process conflict} (disagreement on how to complete the task). The former and latter represent gaps in the SMM, the intermediate reflects an affective dimension
We identified a canonical conflict resolution protocol, where team members identify, acknowledge and scope a conflict, then iteratively apply conflict resolution strategies and evaluate their success. The rubric incorporates the distinction of types of conflict into the various stages of the protocol; in scoping a conflict, a high-performing subject would ascertain its type, then (s)he would apply conflict resolution strategies appropriate for this type, and so forth. The rubric we derived also drew on validated coding schemes, such as \cite{hao2025automated,ouyang2023artificial}. 

Our project management model focuses on two phases of task completion \cite{marks2001, salas2005big5}. The first phase is defining a {\em task model}, which establishes shared understanding of the problem and goals, and a {\em team model}, which delineates individual roles and responsibilities. These two components establish and sustain the shared mental model, and are the main focus of the project management task. These processes are the basis for the first two dimensions of the rubric—{\em goal setting and shared task definition}, and {\em collaborative planning and role definition}. In the second phase, project management functions shift primarily to monitoring progress toward the goal, which is captured by the final dimension {\em mutual status monitoring}. During the collaboration, the team continuously moves between these phases. This framework assumes a {\em shared leadership model}, where responsibility for initiating these phases and associated functions is distributed among the team members \cite{carson2007sharedleadership}.

Two collaborative task frameworks have been shown to effectively elicit collaboration skill, consensus-building and negotiation tasks\cite{haocui2025}. Consensus-building tasks require participants to weigh different viewpoints, opinions or arguments, and then converge on a mutually agreeable decision. Negotiation tasks require participants with varying backgrounds, access to information or conflicting goals to co-develop a solution that satisfies the group goals. Our evaluation used a debate task, and a science experiment task aligned with these frameworks.  

\subsection{Creativity}
Reflecting common modern frameworks \cite{cuestahincapie2025evaluating}, our conceptual model defines creativity as the process of developing outputs—be they products, solutions, or ideas—characterized by being both original and useful. These frameworks typically emphasize a core set of cognitive skills that align with the iterative stages of creative thinking: identifying the challenge, generating and exploring ideas, and evaluating the most effective approach.\cite{brandt2023measuringcreative,SaidMetwaly2017MeasuringCreativity}. Our rubric focused on the \emph{Generating Ideas} and \emph{Evaluating Ideas} steps, and a key supporting disposition, \emph{Building on Ideas}. 

Generating Ideas encompasses the concept of divergent thinking—the cognitive capacity to generate multiple distinct solutions to an open-ended problem. This is widely considered to be the foundation of creativity and innovation, as beginning with a diverse set of initial ideas increases the probability of identifying original, high-utility solutions \cite{aschauer2022scientificCreativity}. The rubric evaluates three measures of idea generation: {\em fluidity}, which gauges the number and variety of distinct ideas; {\em originality}, which measures the degree of novelty; and {\em quality}, which assesses whether the ideas are feasible and address the core challenge\cite{heard2025ACERcreative,acar2025CreativityAssessment,patterson2025CAP}.

While divergent thinking produces a range of ideas, convergent thinking is crucial for determining which concepts possess the highest probability of success \cite{aschauer2022scientificCreativity}. This serves as the foundation of Evaluating Ideas – the cognitive process of assessing, refining, and choosing the most appropriate concept based on specific constraints and criteria. The rubric for this sub-skill considers two components: \emph{elaborating}, the skill of clearly defining an idea's potential by adding specifications and detail \cite{brandt2023measuringcreative,OECD2019creativeAndCritical}; and \emph{selecting}, which assesses whether the subject chooses solutions for implementation that are both original and high-quality\cite{heard2025ACERcreative}.

Finally, our rubric includes Building on Ideas, defined as the ability to effectively incorporate and expand on others' contributions. This skill requires participants to recognize connections and to integrate personal ideas with those of collaborators to produce novel concepts \cite{heard2025ACERcreative}. During the evaluation phase, groups critically assess and refine their proposals, leading to greater originality and improved problem-solving \cite{brandt2023measuringcreative,OECD2019creativeAndCritical}. 

Research suggests that creative thinking assessments are most effective when they are open-ended—allowing participants to fully demonstrate divergent thinking \cite{abdulla2018creative}—and mirror authentic problem—solving scenarios \cite{OECD2019creativeAndCritical}. Following these guidelines, we developed an innovation challenge framework where participants collaborate with an agent to design a creative solution to a loosely defined problem. Examples include developing a skill-sharing system to connect local residents or designing a zero-waste community festival.

\subsection{Critical thinking}

Building on the foundational work of \cite{facione1990delphi,Ennis1991criticalTA, paul2006miniature}, our conceptual model defines critical thinking as a self-regulated cognitive process involving the interpretation, analysis, evaluation, and inference of information, driven by a disposition toward truth-seeking and open-mindedness. Thus, critical thinking requires integrating a combination of cognitive skills. Our evaluation focuses on two of these: {\em Interpret and Analyze,} and {\em Evaluate and Judge}.

The first category, Interpret and Analyze, encompasses the ability to effectively distill the substance of various data, experiences, or beliefs, while mapping reasoning to identify inferential relationships within arguments \cite{heard2025ACERcreative}. Together, these processes {\em clarify intent} and {\em outline arguments} by parsing the meaning and logical structure of complex information.

The second category, Evaluate and Judge, involves appraising the credibility of statements and the weight of inferential relationships \cite{dwyer2014integrated}. Together, these processes allow participants to {\em assess information} by verifying source reliability—including expertise and potential bias—and to {\em assess reasoning} by determining the soundness of arguments and detecting logical errors \cite{heard2025ACERcreative, OECD2019creativeAndCritical}.

Effective AI tool usage is integrated into our assessment of each sub-skill, requiring participants to demonstrate strategic engagement alongside rigorous fact-checking and the evaluation of AI-generated content for potential bias \cite{oecd2026digital}.

Research indicates that critical thinking is best elicited through ``ill-structured'' or ambiguous materials—such as those containing contradictory information—rather than well-structured, cogent texts \cite{liu2016tough}. To maximize this effect, assessments should emulate authentic scenarios, a strategy that has been shown to improve both validity and student engagement \cite{halpern1998teaching}. Applying this framework, we developed a ``Chief Editor'' task in which participants analyze a flawed article draft and determine whether to recommend it for publication.

\section{Collaboration Rubrics and Tasks}

\begin{table}[H]
\scriptsize
\centering
\caption{Conflict Resolution rubrics}
\renewcommand{\arraystretch}{1.65} 
\begin{tabular}{| p{0.11\textwidth} | p{0.19\textwidth} | p{0.19\textwidth} | p{0.20\textwidth} | p{0.20\textwidth} |}
\hline
\textbf{Category} & \textbf{Beginning (1)} & \textbf{Emerging (2)} & \textbf{Developing (3)} & \textbf{Demonstrating (4)} \\ \hline

\textbf{Overall} & 
Fails to identify conflicts, ignores them, or behaves in ways that create or escalate conflict. Their actions are often counter-productive to the team's goals and social climate. & 
Recognizes the presence of conflict but struggles to address it constructively or consistently. May lack the skills to accurately scope the issue or apply effective resolution strategies. & 
Effectively responds to and resolves acknowledged conflicts. Can follow the resolution procedure and apply appropriate strategies but may not always investigate the conflict's deeper dimensions. & 
Proactively identifies and resolves conflicts, investigating root causes, validating that the conflict has been resolved, and strengthening team cohesion. Not only fixes the problem but also improves the team's shared understanding and psychological safety. \\ \hline

\textbf{Conflict Identification (Cognitive)} & 
Fails to identify or ignores conflict. Proceeds with the task as if no disagreement exists, even when it is overt. & 
Acknowledges conflict only when prompted. May show non-verbal signs of disagreement but does not articulate the conflict without being asked directly. & 
Clearly identifies overt conflicts. Acknowledges when a gap in understanding or agreement has been expressed by others. & 
Proactively surfaces potential gaps. Anticipates and points out discrepancies before they become overt conflicts. \\ \hline

\textbf{Scoping and Analysis (Cognitive/Executive)} & 
Mischaracterizes the conflict. Frames a task disagreement as a personal attack, escalating it into a relationship conflict. & 
Oversimplifies the conflict. Describes the disagreement in binary terms (e.g., 'right vs. wrong') without exploring the nuances. & 
Scopes the 'width' of the conflict. Focuses on the surface-level disagreement to understand what is being debated. & 
Investigates the conflict's 'depth' and 'type'. Asks questions to distinguish between a task, process, or relationship conflict and to uncover underlying assumptions or values. \\ \hline

\textbf{Strategy Application (Executive)} & 
Applies detrimental strategies. Employs tactics that worsen the conflict, such as stonewalling, dominating the conversation, or making personal attacks. & 
Applies a limited or mismatched strategy. Uses a single default strategy (e.g., always avoiding or always compromising) regardless of the conflict type. & 
Applies standard resolution practices. Relies on common procedures like voting or compromise, which are generally effective but may not be optimal for the specific situation. & 
Applies adaptive resolution practices. Suggests a resolution strategy tailored to the conflict type (e.g., debate for a task conflict, negotiation for a process conflict). \\ \hline

\textbf{Monitoring and Regulation (Socio-Emotional)} & 
Creates a hostile environment. Uses sarcasm, personal insults, or blame, destroying trust and making others unwilling to contribute. & 
Shows signs of negative emotional response. Expresses frustration, impatience, or annoyance through tone or words, which can reduce psychological safety. & 
Maintains a respectful tone. Engages in the conflict resolution process without resorting to personal criticism or dismissive language. & 
Fosters psychological safety. Actively manages the emotional tone, de-escalates tension, and checks in with teammates to ensure they feel heard and respected. \\ \hline
\end{tabular}
\label{tab:conflict_rubric}
\end{table}

\begin{table}[H]
\scriptsize
\centering
\caption{Project Management rubrics}
\renewcommand{\arraystretch}{1.5}
\begin{tabular}{| p{0.12\textwidth} | p{0.19\textwidth} | p{0.19\textwidth} | p{0.20\textwidth} | p{0.20\textwidth} |}
\hline
\textbf{Category} & \textbf{Beginning (1)} & \textbf{Emerging (2)} & \textbf{Developing (3)} & \textbf{Demonstrating (4)} \\ \hline

\textbf{Overall Competence} & 
Antagonistic/apathetic engagement. Ignores or dismisses the team's plan.  Does not provide status updates or ignores request for status updates. Doesn't consider current status of work/progress in the plan. 'Goes rogue' or continues working on their own idea, while ignoring team members. OR Low effort participation. Relies on others for task completion/planning. Participation is limited to simple replies that do not further or add value to the plan. Does not participate in any form of task completion or planning. & 
Reactive participation. Follows the plan, but does not actively participate in the construction. Relies on teammates to define the plan and assign roles. Steps in before there was a chance to negotiate a plan and insists on their individual task execution even when the task is moving forward otherwise. Gives status update if prompted. & 
Active participation. Contributes to the construction of the plan and shares leadership with group members, but lacks explanations or matching it to the task status/ team members. Initiates task allocation, but not based on current task status or skill-match. Asks clarifying questions about goals and roles, helping to refine the plan. Contributes practical ideas and suggestions. Acts decisively when things are stuck or off-track but doesn't explain that explicitly. Contributes to shared awareness of task progress by openly sharing their own task status, progress, and blockers without needing to be prompted. & 
Leads the co-construction and adaptive planning proactively. Facilitates discussion to ensure a shared understanding of the task and that all voices are heard - based on task allocation and current task status. Answers clarifying questions about goals and roles, displays overarching understanding. Acts decisively when things appear to be stuck or off-track and is able to adapt/change the plan if needed and explains why. Prompts team members to share the status of their tasks and supporting evidence. Analyzes evidence and provides constructive feedback. \\ \hline

\textbf{Goal Setting \& Shared Task Definition} & 
Ignores, dismisses, or undermines the team's goal-setting process. Creates confusion during the team's goal setting discussion. & 
Passively accepts the team's goals. Follows the discussion but does not actively participate in constructing or defining the team's goals. & 
Actively contributes to the discussion and definition of team's goals. Asks clarifying questions to help refine the team's shared understanding of goals or tasks. & 
Initiates and facilitates the co-construction of clear, shared, and measurable goals. Answers clarifying questions about goals and task requirements. Displays overarching understanding of task goals. \\ \hline

\textbf{Collaborative Planning \& Role Definition} & 
Ignores or dismisses the team's planning and/or has a low effort participation in task planning process. Ignores the team's agreed-upon plan or roles, insisting on their own planning without grounding it with the team. Or participation is limited to simple replies that do not further or add value to the plan. Does not participate in any form of task completion or planning. Does nothing if the team appears to be stuck. & 
Relies on others for defining the plan. Follows the plan, but does not actively participate in the construction. Relies on teammates to define the plan and assign roles. Steps in before there was a chance to negotiate a team plan and insists on their individual task execution even when the task is moving forward otherwise. If the team appears to be stuck, either follows others direction (e.g. not taking initiatives or proposing solutions) or only suggests arbitrary task division. & 
Actively contributes practical ideas to the planning, but does not lead the discussion, and/or does not display adaptivity to the task's planning. Asks clarifying questions about goals and roles, helping to refine the plan. Contributes practical ideas and suggestions. Initiates task allocation, but not based on the current task status or skill-match. Does not modify the current plan based on status. If the team appears to be stuck, proposes a solution for moving forward, but does not explain the reasoning behind the solution or does not provide an explanation of the logjam. & 
Leads the co-construction and adaptive planning proactively. Facilitates discussion to ensure a shared understanding and that all voices are heard for task planning and task allocation. Answers clarifying questions about roles, displays overarching understanding. Decisively modifies the plan if and when needed, with inputs/agreement from the team or makes an individual decision but is able to explain why. If the team appears to be stuck, provides an explanation of the logjam, a solution and the reason why a particular approach can help move things forward better than an alternative. \\ \hline

\textbf{Mutual Status Monitoring} & 
Limits or hinders team's shared awareness about task progress by not providing status updates when prompted or ignores request for status update. & 
Has limited contribution to team's shared awareness and/or seems to have limited awareness of task progress of the team. Shares status update if prompted by team members, but is not actively contributing to the overall team's shared awareness of team's progress. & 
Contributes to shared awareness of task progress by openly sharing their own task status, progress, and blockers without needing to be prompted. Does not necessarily inquire about the team members' task status. & 
Creates shared awareness of task progress by sharing their own task status and prompts team members to share theirs. Prompts team members to share the status of their tasks and supporting evidence. Analyzes evidence and provides constructive feedback. \\ \hline
\end{tabular}
\label{tab:project_management_rubric}
\end{table}

\begin{table}[H]
\caption{Collaboration skill tasks}
\label{tab:si_tasks}
\centering
\scriptsize
\renewcommand{\arraystretch}{1.5}
\begin{tabular}{|p{0.5\linewidth}|p{0.5\linewidth}|}
\hline
\textbf{Task 1: Science Experiment} & \textbf{Task 2: Debate} \\
\hline
The school grant committee approved your science experiment on the Truth in advertisement. \newline\newline
Now you need to translate your vision to a research plan, and apply the initial stages of the Scientific Method. \newline\newline
The goal is to explore the topic you selected: ``Truth in advertisement: does brand A paper towel really absorb more than brand B?'' \newline\newline
To address this question, before designing the actual experiment, you need, as a group, to provide the following: \newline\newline
\textbf{Question:} Formulate a testable question based on the provided scenario. \newline
\textbf{Hypothesis:} Develop a hypothesis that offers a potential answer to your question. \newline
\textbf{Experiment:} Design an experiment to test your hypothesis. This should include: a comprehensive procedure and a list of all materials required. \newline\newline
Good luck experts! 
& 
Your group is an expert panel that will discuss the following question: \newline\newline
Should the government regulate social media content? \newline\newline
As a team you need to: \newline\newline
\textbf{1. Choose Your Viewpoint:} \newline
$\bullet$ In Favor \newline
$\bullet$ Against \newline\newline
\textbf{2. Develop Arguments:} \newline
Brainstorm arguments that support your chosen viewpoint. Consider the potential benefits and drawbacks. \newline
Aim to decide on at least three strong arguments. \newline
Each member should contribute at least one argument. \newline\newline
\textbf{3. Prepare an Opening Statement:} Work together to write a short opening statement that clearly presents your panel's viewpoint and summarizes your three main arguments. \newline\newline
Remember to collaborate, be persuasive, and support your claims with evidence \newline\newline
Good luck, experts \\
\hline
\end{tabular}
\end{table}

\section{Creativity Rubrics and Tasks}

\begin{table}[H]
\scriptsize
\centering
\caption{Generating Ideas rubrics}
\renewcommand{\arraystretch}{1.65} 
\begin{tabular}{| p{0.10\textwidth} | p{0.20\textwidth} | p{0.20\textwidth} | p{0.20\textwidth} | p{0.20\textwidth} |}
\hline
\textbf{Category} & \textbf{Dormant (1)} & \textbf{Emerging (2)} & \textbf{Demonstrating (3)} & \textbf{Excelling (4)} \\ \hline
\textbf{1. Fluidity} Produces large variety, distinct ideas to consider
 & Produces a well below average number of thematically distinct ideas. OR Ideas cover a well below average range of conceptual categories for the number of ideas. & Produces a below average number of thematically distinct ideas. OR Ideas cover a below average range conceptual categories for the number of ideas. & Produces an average or above average number of thematically distinct ideas. AND Ideas cover an average or above average range of conceptual categories for the number of ideas. & Produces a well above average number of  thematically distinct ideas. AND Ideas cover a well above average range of conceptual categories for the number of ideas. \\ \hline

\textbf{Behaviors / Indicators} & 
*Adopts a singular perspective

*Does not shift viewpoints when prompted.

*Repeats essentially the same ideas with small iterations.

*Ideas are limited to a single category & 
*Considers a limited number of related perspectives

*Makes predictable shifts in viewpoint when prompted 

*Suggests small variations on common solutions or approaches 

*Ideas fall into related categories &
\multicolumn{2}{p{0.40\textwidth}|}{
*Considers a range of perspectives  and approaches

*Adjusts constraints to generate new ideas

*Ideas include distinct themes (e.g., bring your own bottle, water fountains)

*Ideas correspond to distinct conceptual categories

*Explores unexpected viewpoints and approaches

*Redefines problem boundaries to open new conceptual categories} \\ \hline

\textbf{2. Originality}
Produces novel ideas  (relative to user background)  & Produces well-known solutions or highly predictable approaches.
 & Produces slight variations on common solutions or approaches. & 
Produces unconventional solutions or approaches. & 
Produces surprising or novel solutions or approaches. \\ \hline

\textbf{Behaviors / Indicators} &
*Does not consider combining ideas

*Presents only familiar, or conventional solutions &
*Only combines obviously related ideas

*Makes limited, superficial adaptations &
*Combines ideas in unconventional ways or that don't seem obviously related

*Presents ideas that demonstrate a clear departure from common approaches &
*Reframes ideas and tasks in surprising ways to bring new interpretations of what is possible.

*Presents ideas that are exceptionally unique or integrate entirely unexpected elements into known approaches. \\ \hline

\textbf{3. Quality}
Produce feasible ideas that address the problem or challenge & Suggestions are clearly irrelevant (e.g., using a banana as a hammer), infeasible (e.g., require vast resources) or by ignoring obvious, practical constraints (e.g., time, space). &
Suggestions address only a portion of the problem or surface-level symptoms rather than the core challenge, require significant resources, or only partially satisfy constraints. &
Ideas are both relevant and feasible. They directly address the core challenge and satisfy the main constraints. &
In addition to being relevant and feasible, ideas meet all criteria and are adaptable to changing conditions or impediments. \\ \hline

\textbf{Behaviors / Indicators} &
*Ideas require unavailable, cutting edge technology &
*Ideas require extensive budget or time that significantly outweigh any potential benefit.

*Ideas will only work in limited conditions&
*Ideas that are practical, meet the main criteria

*Ideas are likely to work for most conditions

*Pivots base ideas to meet new constraints or respond to impediments &
*Potential benefit greatly exceeds the required resources \\ \hline

\end{tabular}
\end{table}

\begin{table}[H]
\scriptsize
\centering
\caption{Evaluating Ideas rubrics}
\renewcommand{\arraystretch}{1.65} 
\begin{tabular}{| p{0.11\textwidth} | p{0.19\textwidth} | p{0.19\textwidth} | p{0.20\textwidth} | p{0.20\textwidth} |}
\hline
\textbf{Category} & \textbf{Dormant (1)} & \textbf{Emerging (2)} & \textbf{Demonstrating (3)} & \textbf{Excelling (4)} \\ \hline
\textbf{1. Elaborating} Develops ideas by adding adding detail and depth & Provides little to no detail or description of the idea. &
Provides a surface-level description of ideas, but does not explain how it would work. &
Provides details beyond surface-level specs that describe how the idea would work. However, it lacks sufficient detail to support prototyping. &
Presents substantive explanations of ideas including functionality. Provides enough detail to prototype the ideas. \\ \hline

\textbf{Behaviors / Indicators} &
*Ideas are skeletal or use vague terms (e.g., "an app," "a tool").&
*Describes what the idea is but not how it functions.&
*Explains the logic, workflow, or "engine" of the idea. &
*Uses specific technical or functional parameters (e.g., "a geo-location-based API," "a weighted mallet"). \\ \hline

\textbf{2. Selecting} Choose novel, high quality solutions to implement
 & Chooses to implement ideas that fail to address the core problem, are infeasible, or do not meet practical key criteria. &
Chooses to implement  ideas that only partially address the core problem, have limited feasibility, or only satisfies some practical constraints. Favors novelty at the expense of utility (or vice versa). Relies on surface-level criteria for selection. &
Chooses to implement  feasible ideas that address the core problem and meet practical constraints. Balances utility, and novelty with other key criteria when selecting ideas. &
In addition to being feasible, addressing the core problem, meeting practical constraints, and balancing utility and novelty,  chooses to implement ideas that are adaptable to changing conditions or criteria. \\ \hline

\textbf{Behaviors / Indicators} &
*Chooses based on personal preferences

*Only considers surface level characteristics

*Ignores or misses obvious failure points, gaps, or other challenges

*Applies flawed criteria (e.g., sunk costs) &
*Selects a few criteria for evaluating the idea, ignoring others

*Selects ideas that address obvious symptoms but miss the underlying challenges.

*Selection favors "easy wins" or familiar territory

*Focuses solely on novelty or utility, or another chosen criteria &

*Identifies obvious failure points or challenges

*Select demonstrates consideration of logical "trade-offs" &

*Proactively anticipates subtle failure points or challenges.

*Identifies risks and plans for iteration \\ \hline

\end{tabular}
\end{table}

\begin{table}[H]
\scriptsize
\centering
\caption{Building on Ideas rubrics}
\renewcommand{\arraystretch}{1.65} 
\begin{tabular}{| p{0.11\textwidth} | p{0.19\textwidth} | p{0.19\textwidth} | p{0.20\textwidth} | p{0.20\textwidth} |}
\hline
\textbf{Category} & \textbf{Dormant (1)} & \textbf{Emerging (2)} & \textbf{Demonstrating (3)} & \textbf{Excelling (4)} \\ \hline
\textbf{Overall Competence} The process of transforming and integrating one's own ideas with other’s contributions to broaden the range of possibilities. & Ignores or dismisses others’ suggestions to improve or adjust own ideas.&
Focuses on adjusting one's own ideas rather than building on or incorporating others. Makes small, "safe" modifications based on others suggestions.&
Actively adapts and improves concepts by incorporating or expanding on suggestions or ideas provided by others.&
Seamlessly weaves together own ideas with the ideas of others to create a stronger, more robust concept. \\ \hline

\textbf{Behaviors / Indicators} &
*Ideas are skeletal or use vague terms (e.g., "an app," "a tool") & 
*Describes what the idea is but not how it functions &
*Explains the logic, workflow, or "engine" of the idea & 
*Uses specific technical or functional parameters (e.g., "a geolocation-based API," "a weighted mallet")  \\ \hline

\end{tabular}
\end{table}

\begin{table}[H]
\scriptsize
\centering
\caption{Creativity tasks}
\renewcommand{\arraystretch}{1.65} 
\begin{tabular}{| p{0.08\textwidth} | p{0.09\textwidth} | p{0.53\textwidth} | p{0.18\textwidth}|}
\hline
\textbf{Theme} & \textbf{Task} & \textbf{Description} & \textbf{Submission} \\
\hline
Community / Society & 
Skill-share innovation challenge &
You are competing in an innovation challenge to design a skill-sharing system for apartment residents. Work with your partner to explore community needs and develop a solution that is both original and practical. Consider the following in your design:
\begin{enumerate}
\item Map Possibilities: List various methods to manage skill sharing between residents. 
\item Develop Concepts: Integrate and expand ideas from your list into a few detailed system concepts.
\item Refine: Choose your best concept and explain how your system will create connections between residents to support each other.  Submit your final design.
\end{enumerate} &
Outline your skill-sharing system and explain how it supports resident connections. \\
\hline
Science / STEM &
Zero-waste Festival &
You are entering a state-wide contest to design a community Earth Day festival with the lowest possible waste output. Work with your partner to rethink traditional event elements like food and activities to minimize or eliminate garbage or recycling. Your conversation should follow these three stages:
\begin{enumerate}
\item Map Possibilities: List ways to reimagine festival elements that reduce or element waste. 
\item Develop Concepts: Integrate and expand ideas from your list into a few detailed concepts for sustainable festivals.
\item Refine: Choose your best concept and explain how your festival concept minimizes waste without sacrificing fun. Submit your final design.
\end{enumerate} &
Describe your festival concept, outlining the elements and explaining how it meets your zero-waste goals. \\
\hline
\end{tabular}
\end{table}

\clearpage
\section{Critical Thinking Rubrics and Tasks}

\begin{table}[H]
\ssmall
\centering
\caption{Critical thinking Interpret and Analyze rubrics}
\begin{tabular}{| p{0.12\textwidth} | p{0.2\textwidth} | p{0.2\textwidth} | p{0.2\textwidth} | p{0.15\textwidth} |}
\hline
\textbf{Category} & \textbf{Dormant (1)} & \textbf{Emerging (2)} & \textbf{Demonstrating (3)} & \textbf{Excelling (4)} \\ \hline
{\bf Overall}.

Identify core components, determine the logical structure, and clarify the intended meaning of complex information, observations, or arguments. &
Treats the text as a flat narrative rather than a structured argument. Confuses narrative ``filler'' text with actual claims or core concepts. Fully relies on others or AI tools for interpretation or analysis. & Interprets statements rigidly and in isolation without considering the underlying context. 

Identifies that an argument exists but mislabels the components, such as confusing a premise with a conclusion or facts with subjective observations.

Uses the tool only for simple information retrieval, or without providing context. &
Clarifies ambiguous terms in context. 

Accurately maps the structure of an argument (i.e., premises to conclusions). 

Uses AI to retrieve specific, relevant information within a defined context. &
Considers intent along with context to determine deeper meaning. 

Identifies missing or omitted information, unstated assumptions, equivocations, or gaps in reasoning. 

Strategically uses AI to uncover omitted details or identify statements with multiple underlying meanings.
\\ \hline
{\bf 1. Clarifying Intent}.

Separating core claims and evidence from ``filler'' text to understand meaning &
Misidentifies core components by treating narrative filler, anecdotes, or introductory remarks as informational content (e.g., primary claims or evidence). &
Distinguishes informational content from narrative filler but miscategorizes the role of statements in context. &
Accurately categorizes informational statements and determines their role. context. &
Consider intent along with context when determining meaning.
\\ \hline
{\bf 2. Outlining Arguments}.

Mapping connections between premises, conclusions, and supporting evidence. &
Treats the text as flat information , failing to detect when an argument is being made or that statements are playing a logical role. &
Identifies that an argument exists but mislabels the components, misidentifies the role of statements, inaccurately maps the relationship between premises, evidence, and conclusions, or overlooks obvious flaws in reasoning. &
Deconstructs and maps the structure of the argument accurately. Identifies explicit gaps in reasoning or obvious unstated assumptions. &
Identifies ``hidden'' assumptions, subtle gaps in the logic, and whether any critical evidence is missing or omitted. 
\\ \hline
{\bf 3. AI-Supported Exploration}.

Using AI tools to retrieve and clarify information &
Attempts to fully outsource interpretation or analysis to the AI tool or collaborators. &
Uses the tool only for simple information retrieval. Does not provide specific details or context. &
Use the tool to find specific, relevant information. Provides the AI with specific context to get better results &
Use the tool to identify missing or omitted information, and identify terms or statements that have multiple or subtle underlying meaning. 
\\ \hline
{\bf Behaviours / Indicators} &
* Does not distinguish narrative elements of the text from key statements, such as claims (i.e., premises, conclusions), evidence (e.g., facts, data), or concepts. 

* Confuses narrative elements or ``filler'' text with actual claims and evidence or other core information.

* Accepts ambiguous terms, jargon, or statements without seeking clarification

* Does not detect that an argument is being made. 

* Does not identify when a statement is playing a logical role

* Attempts to fully outsource reading comprehension or structural analysis to the AI tool 

* Relies entirely on the others to explain the logic. &
* Distinguishes core information from filler by identifying statements with informational content. 

* Confuses facts with expert opinions or value judgments.

* Interprets statements rigidly without considering the underlying context

* Mislabels the components of an argument (e.g., confusing a premise with a conclusion, facts with subjective observations)

* Inaccurately maps premises or supporting evidence to conclusions

* Fails to identify obvious missing assumptions or gaps in reasoning

* Uses the AI tool for basic information retrieval (e.g., generic summarization or dictionary definitions) without providing specific details or context. &
* Distinguishes facts from expert opinions of value judgments. 

* Clarifies ambiguous terms or statements in context 

* Identify the role of specific terms or phrases in a text

* Consider context when interpreting information   

* Breaks an argument down into its constituent parts: premises and conclusions 

* Accurately trace the relationship between parts of an argument

* Identifies obvious missing assumptions or gaps in reasoning 

* Uses the AI tool to fact-check or retrieve relevant information, definitions, or summaries. Provides relevant context.

* Identifies the relevant information from AI response. &
* Consider intent along with context when interpreting information or clarifying terms

* Strategically queries the AI tool to find specific omitted information, reveal unstated assumptions, or expose equivocations \\
\hline
\end{tabular}
\end{table}

\begin{table}[H]
\ssmall
\centering
\caption{Critical thinking Evaluate and Judge rubrics}
\begin{tabular}{| p{0.12\textwidth} | p{0.18\textwidth} | p{0.2\textwidth} | p{0.2\textwidth} | p{0.17\textwidth} |}
\hline
\textbf{Category} & \textbf{Dormant (1)} & \textbf{Emerging (2)} & \textbf{Demonstrating (3)} & \textbf{Excelling (4)} \\ \hline
{\bf Overall}.
Assess the credibility of sources and the logical validity of claims by applying rigorous intellectual standards and identifying potential fallacies or biases. &
Evaluates arguments based primarily on personal agreement with the conclusion. 

Selects evidence purely to support pre-existing beliefs. 

Attempts to fully outsource the task to the AI tool. &
Evaluates sources based on surface traits or ‘common sense’. 

Readily accepts common fallacies (e.g., emotional appeals or logical traps). 

Uses the AI tool for surface level fact checking. Does not critically engage with the output. &
Evaluates sources and evidence using established criteria. 

Identifies logical fallacies or errors in reasoning. 

Actively uses verified tool outputs to cross-reference claims, or identify contradictions and fallacies. &
Evaluates quality of sources and evidence with respect to specific claims.

Explains why an argument is flawed (e.g., invalid, unsound, fallacious).

Uses AI to stress-test arguments.
\\ \hline
{\bf 1. Assessing Information}.

Evaluating whether evidence is reliable using established criteria &
Selects information based on pre-existing beliefs. Dismisses opposing evidence without analysis. &
Evaluates sources to evidence based on surface characteristics or 'common sense'. Accepts flawed or incomplete information. &
Systematically evaluates sources using established criteria (relevance, accuracy, currency) and correctly identifies the specific evidence required to validate an argument. &
Evaluates evidence with respect to specific claims. Considers subtle issues, such as conflicts or interests or subtle misrepresentation. Seeks out evidence to stress test own position.
\\ \hline
{\bf 2. Assessing reasoning}.

Judging the integrity of an argument by identifying fallacies and whether the premises support the conclusion. &
Evaluates article reasoning primarily on whether or not they agree with the conclusion. 
Does not identify obvious logical flaws. &
Identifies obvious logical flaws but fails to diagnose the specific error. Labels arguments as ``wrong'' without providing evidentiary reasoning. Accepts rhetoric or common logical fallacies. &
Identifies the invalid logic or unsound of arguments. Correctly identifies common logical fallacies or errors in reasoning. &
Explains exactly why an argument is valid and sound, or not. Explains why a fallacy invalidates the logic and how to mitigate the issue. 
\\ \hline
{\bf 3. AI-Supported Verification}.

Using AI tools to cross-reference claims, verify details, and find missing or misleading information. &
Attempts to fully outsource evaluation or judgement to the AI tool or collaborators. &
Uses the AI tool as a surface-level fact checker (e.g., checking a single statistic). Does not critically engage with or apply  the output. &
Uses AI tools to fact-check or retrieve relevant information. Uses those output to identify contradictions, misused or misinterpreted evidence, or flawed reasoning. &
Uses the tool ``adversarially'' to stress-test an argument by searching for counter-evidence, verifying methodology, or source background 
\\ \hline
{\bf Behaviours / Indicators} &
* Filters or selects sources or information purely to support pre-existing beliefs.

* Dismisses opposing views or counter-evidence without analysis or engagement.

* Labels an argument as 'wrong' or 'illogical' based primarily on personal disagreement with the conclusion.

* Dismisses opposing views  without analysis or engagement.

* Attempts to fully outsource subjective evaluation or decision-making to the AI tool.

* Blindly accepts claims without utilizing the tool for verification. &
* Evaluates sources or evidence based on surface characteristics, (e.g., tone, popularity, or perceived authority) or 'common sense'. 

* Accept evidence that is incomplete, irrelevant, or misleading.

* Accepts common fallacies (e.g., hasty generalization, false dichotomy) 

* Labels an argument as wrong or illogical without providing supporting evidence or reasoning

* Accepts persuasive, flawed rhetoric (e.g., emotional appeals, appeals to authority) 

* Quotes AI outputs without critical evaluation, or intellectual oversight.

* Does not  leverage AI outputs to evaluate the argument's logic, or does so inaccurately or ineffectively. &
* Correctly assesses sources using standard markers (e.g., expertise, currency, relevance). 

* Misses nuanced issues in sources or evidence (e.g., conflicts of interest, methodological issues).

* Identifies evidence needed to assess an argument. 

* Does not actively seek out disconfirming evidence to stress-test their own position.

* Explicitly identifies or names logical fallacies (e.g., correlation vs causation, circular reasoning). 

* Actively uses the AI tool to cross-reference claims and the source evidence. 

* Uses the verified outputs from the AI tool not just to check facts, but to accurately identify and articulate explicit logical fallacies or direct contradictions within the text. &
* Distinguishes between factual accuracy and misleading presentation. 

* Evaluates the source quality relative to the specific claim (e.g., identifying selection bias, relevance, conflicts of interest).

* Seeks opposing evidence or counter-arguments that could invalidate their personal position and claims

* Explains exactly how or why a fallacy invalidates an argument.

* Prompts the AI tool to verify specific methodological details, or retrieve evidence contradicting claims with respect to a specific context

* Uses the AI tool adversarially to stress-test the own argument. 
\\ \hline
\end{tabular}
\end{table}

\begin{table}[H]
\scriptsize
\centering
\caption{Critical Thinking Editorial Review tasks}
\renewcommand{\arraystretch}{1.65} 
\begin{tabular}{| p{0.08\textwidth} | p{0.26\textwidth} | p{0.58\textwidth}|}
\hline
\textbf{Theme} & \textbf{Description} & \textbf{Editorial Draft} \\
\hline
Community / Society & 
You are the lead Technology \& Society Editor at a high-stakes news outlet. A journalist submitted an article on the effect of social media on teen mental health. You must decide whether or not to publish it. Your task is to:
\begin{enumerate}
\item Audit for Flaws: Analyze the article to identify \"red flags\" and use your AI research assistant to fact-check the author’s claims and sources.
\item Evaluate \& Decide: Meet with the journalist to discuss their research and determine whether the article meets the standards for publication.
\end{enumerate}
You have access to an AI tool for fact-checking, summarization, information retrieval, and more. Simply address ``AI tool'' in the chat to invoke it.
&
{\bf The Digital Panacea: Why We Should Stop Policing Teen Social Media Use}.

For decades, parents and pediatricians have treated adolescent social media use as a crisis. However, recent clinical reviews suggest it is time to abandon outdated policing frameworks in favor of digital autonomy. With 90\% of 13 to 17-year-olds using YouTube and 63\% on TikTok, the digital landscape is not a threat, but a primary healthcare provider.

Research shows that teens are highly discerning consumers of online medical content, independently fact-checking the health information they learn. Because adolescents actively use these platforms to build coping strategies and find safe environments to discuss mental health, clinical oversight is largely redundant. If adolescents are successfully self-diagnosing and building community support systems online, enforcing ``family media plans'' only disrupts their natural developmental identity formation.

Furthermore, the fear that screens are ``crowding out'' physical activities is fundamentally misunderstood. Studies confirm that active participation on social media is associated with reduced loneliness. Therefore, spending hours actively scrolling and engaging online is socially and developmentally equivalent to in-person connection, rendering ``screen time limits'' obsolete.

Finally, the data is clear on mental health outcomes: hospitalized adolescents frequently use social media as a means of emotional distraction from negative thoughts. By providing an escape from mental illness symptoms, unrestricted social media access effectively prevents the severe depressive episodes that lead to hospitalization. It is time we recognize social media not as a vice to be managed, but as a self-regulating medical intervention.
\vspace{0.05in}

{\bf Source article}: Health Benefits of Social Media Use in Adolescents and Young Adults (\url{http://pmc.ncbi.nlm.nih.gov/articles/PMC12356748})
\\ \hline
Science / STEM &
You are the lead Science \& Health Editor at a high-stakes news outlet. A journalist submitted an article draft on the risks of drinking coffee. You must decide whether or not to publish it. Your task is to:
\begin{enumerate}
\item Audit for Flaws: Analyze the article to identify ``red flags'' and use your AI research assistant to fact-check the scientific claims and study data.
\item Evaluate \& Decide: Meet with the journalist to discuss their findings and determine whether the article meets the standards for publication.
\end{enumerate}
You have access to an AI tool for fact-checking, summarization, information retrieval, and more. Simply address ``AI tool'' in the chat to invoke it.
&
{\bf The Bitter Truth: Why It's Time to Reconsider Your Daily Brew}

For decades, coffee lovers have clung to the comforting myth that their morning ritual is harmless. However, a sweeping new clinical analysis of over 9,000 participants from the Hamburg City Health Study has shattered this complacency, exposing the severe cardiovascular toll of our daily brew.

The data is unequivocal: high coffee consumption directly drives a significant spike in LDL-cholesterol. Participants drinking more than four cups a day experienced a sharp, statistically significant increase in this dangerous biomarker. The researchers' regression models meticulously adjusted for age, BMI, smoking, and additives like sugar and milk, proving that the beverage itself is the culprit, not just poor lifestyle choices.

Medical science has long established that elevated LDL-cholesterol is a primary catalyst for major cardiovascular diseases, including heart failure. Therefore, by aggressively elevating LDL levels, heavy coffee consumption systematically accelerates the onset of severe cardiac conditions. The structural mechanism is clear: the bioactive compounds in unfiltered coffee suppress LDL receptor activity, leading to a dangerous extracellular accumulation of cholesterol.

Public health officials must stop giving coffee a free pass. If any pharmaceutical drug caused such a reliable increase in a primary heart failure precursor, it would be immediately pulled from the market. It is time to treat high coffee consumption not as a harmless cultural staple, but as a severe cardiovascular liability.
\vspace{0.05in}

{\bf Source article}: Coffee consumption and associations with blood pressure, LDL-cholesterol and echocardiographic measures in the general population (\url{http://www.nature.com/articles/s41598-023-31857-5})
\\ \hline
\end{tabular}
\end{table}
\end{document}

%% file: macros.tex
\usepackage{xurl}
\usepackage{hyperref}
\usepackage{cleveref}
\usepackage{multirow}
\usepackage{bm}
\usepackage{quoting}
\usepackage{ragged2e}
\usepackage{array}
\usepackage{enumitem}
\usepackage{rotating}
\usepackage{moresize}
\usepackage{float}

\usepackage{booktabs}      
\usepackage{array}         
\usepackage{caption}       
\usepackage{graphicx}
\usepackage{subcaption}
\usepackage{chngcntr}
\usepackage{longtable}
\usepackage{listings}
\usepackage{xcolor}
\usepackage{xspace}

\definecolor{backcolour}{rgb}{0.95,0.95,0.92}
\lstdefinestyle{monospacedgrey}{
    backgroundcolor=\color{backcolour} \& 
    basicstyle=\ttfamily\scriptsize \& 
    numbers=none \& 
    breaklines=true \& 
    breakatwhitespace=true \& 
    postbreak=\mbox{\textcolor{red}{$\hookrightarrow$}\space} \& 
}
\crefformat{footnote}{#2\footnotemark[#1]#3}

\newcolumntype{C}[1]{>{\centering\let\newline\\\arraybackslash\hspace{0pt}}m{#1}}

\newcommand{\takeout}[1]{}

\newcommand{\figref}[1]{Figure~\ref{#1}}

\newcommand{\agbib}[1]{}
\newcommand{\secref}[1]{Section \ref{#1}}
\newcommand{\commentout}[1]{}

\newcommand{\ignore}[1]{}

\newcommand{\puppet}{Executive LLM\xspace}
\newcommand{\AIscorer}{AI Evaluator\xspace}
\newcommand{\nonpuppet}{Independent Agents\xspace}

\newcommand{\skillup}{Vantage\xspace}